\documentclass[12pt,preprint]{aastex}

\newcounter{address}
\newcounter{tableone}
\newcounter{tabletwo}
\newcounter{tablethree}

\newcommand{\arcs}{\unit{arcsec}}
\newcommand{\eg}{\latin{e.g.}}
\newcommand{\ie}{\latin{i.e.}}
\newcommand{\latin}[1]{\textit{#1}}
\newcommand{\unit}[1]{\mathrm{#1}}
\newcommand{\Mpc}{\mathrm {Mpc}}
\newcommand{\hMsun}{h_{70}^{-2}\,\mathrm {M_{\odot}}}

\newcommand{\hMpc}{h_{70}^{-1}\,\mathrm {Mpc}}
\newcommand{\hKpc}{h_{70}^{-1}\,\mathrm {kpc}}

\newcommand{\hGyr}{h_{70}^{-1}\,\unit{Gyr}}
\newcommand{\hhGyr}{h_{70} \,\unit{Gyr}^{-1}}
\newcommand{\hpercentGyr}{h_{70} \, \unit{percent \, per \, Gyr}}
\newcommand{\hnewpercentGyr}{h \, \unit{percent \, per \, Gyr}}
\newcommand{\percent}{\%}
\newcommand{\km}{\mathrm {km}}
\newcommand{\s}{\mathrm {s}}
\newcommand{\di}{\mathrm{d}}
\renewcommand{\mag}{\mathrm {mag}}
\newcommand{\set}[1]{\mathbb{#1}}
\newcommand{\setDDs}{\set{D}_s}
\newcommand{\setDDi}{\set{D}_i}
\newcommand{\setRRs}{\set{R}_s}
\newcommand{\setRRi}{\set{R}_i}

\begin{document}

\title{Galaxy growth by merging in the nearby universe}
\author{
	Tao~Jiang\altaffilmark{\ref{NYU}},
	David~W.~Hogg\altaffilmark{\ref{NYU},\ref{email},\ref{MPI}},
        Michael~R.~Blanton\altaffilmark{\ref{NYU}}
}

\setcounter{address}{1}
\altaffiltext{\theaddress}{\stepcounter{address}\label{NYU} Center for
Cosmology and Particle Physics, Department of Physics, New York
University, 4 Washington Pl, New York, NY 10003}
\altaffiltext{\theaddress}{\stepcounter{address}\label{email} To whom
correspondence should be addressed: \texttt{david.hogg@nyu.edu} }
\altaffiltext{\theaddress}{\stepcounter{address}\label{MPI}
Max-Planck-Institut f\"ur Astronomie, K\"onigstuhl 17, D-69117
Heidelberg, Germany}


\begin{abstract}
We measure the mass growth rate by merging for a wide range of galaxy types.
We present the small-scale ($0.014 < r < 11 \, \hMpc$) projected cross-correlation functions $w(r_p)$ of 
galaxy subsamples from the spectroscopic sample of the NYU VAGC ($5 \times 10^5$ galaxies of 
redshifts $ 0.03 < z < 0.15$) with galaxy subsamples from the SDSS imaging ($4 \times 10^{7}$ galaxies).
We use smooth fits to de-project the two-dimensional functions $w(r_p)$ to obtain smooth 
three-dimensional real-space cross-correlation
functions $\xi(r)$ for each of several spectroscopic subsamples with each of several imaging
subsamples. Because close pairs are expected to merge, the three-space functions 
and dynamical evolution time estimates provide galaxy
accretion rates.
We find that the accretion onto massive blue galaxies and onto red galaxies is
dominated by red companions, and that onto small-mass blue galaxies, red and 
blue galaxies make comparable contributions. We integrate over all 
types of companions and find that at fixed stellar mass, 
the total fractional accretion rates onto red galaxies ($\sim 3$ $\hpercentGyr$) is greater than that 
onto blue galaxies ($\sim 1$ $\hpercentGyr$). 
These rates are almost certainly over-estimates because we have assumed that all close pairs merge
as quickly as the merger time that we used.
One conclusion of this work is that if the total growth of red galaxies from $z=1$ to $z=0$
is mainly due to merging, the merger rates must have been higher in the past.

\end{abstract}



\section{INTRODUCTION}

The galaxy mergers may play an 
important role in the evolution of the galaxies.
In the color-magnitude space, galaxies are separated into two distinct regions:
\textsl{(1)}~the `red sequence': the `early-type', red galaxies ; 
\textsl{(2)}~the `blue cloud' or `blue sequence': the `late-type', blue 
galaxies with strong ongoing star formation \citep{Strateva01a, Blanton03b}.
Some recent studies in the high-redshift ($z \sim 1$) universe find 
that the early-type galaxy population is growing 
over time \citep{bell04a, willmer05a, blanton06a, brown07a,
faber07a, Zhu11a}, which is also found at 
very high ($z \sim 2$) redshift \citep{Daddi05a, Trujillo07a, Longhetti07a, Toft07a, conselice07a, Cimatti08a, vanDokkum08a, Saracco09a}.
In numerical simulations, some studies show us that 
major mergers of intermediate-stellar-mass late-type galaxies maybe play an important role in 
the growth of the intermediate-stellar-mass early type galaxies \citep{barnes96a,naab03a}.
However, the massive early type galaxies may grow in a different way \citep{naab03a}. 
Some recent studies show us that `dry mergers' - the mergers between
early-type galaxies - might play an important role in the 
growth of massive early-type galaxies \citep{bell06b, van Dokkum05a, Masjedi08a}.

There are a lot of studies that estimate the merger rate among galaxies.
These studies can be separated into two general categories.
The studies in the first category count the `pre-merger' close
pairs and convert the `pre-merger' pairs to a merger rate
\citep[\eg,][]{zepf89a, carlberg94a, carlberg00a, carlberg00b, patton97a, patton00a,
van Dokkum99a, lin04a, bell05a, bell06a, kartaltepe07a, Masjedi06a, Masjedi08a,
Patton08a, kitzbichler08a, Bundy09a, De Propris07a, Lin08a, de Ravel09a, De Propris10a, Robaina10a}.
The studies in the second category count the `post-merger' galaxies which have recently experienced at least 
one merger event.
These `post-merger' galaxies are chosen by some observable special properties caused by merging. 
An example might be by star formation indicators of `post-merger' galaxies \citep{quintero04a}
or by morphological signatures caused by merger events
\citep{abraham96a, conselice03a, van Dokkum05a, lotz06a, De Propris07a, Lotz08a, Conselice09a}.

Our work builds on the earlier works of \cite{Masjedi06a, Masjedi08a}, 
which have found previously that luminous red galaxies (LRGs) are growing on average by 
less than 2 $\hpercentGyr$ from merger activity at 
redshifts $0.16 < z < 0.30$ \citep{Masjedi08a}.
In this paper, 
we consider both red and blue galaxies.
We use the previous technique for measuring the close pairs \citep{Masjedi08a} on
NYU VAGC Spectroscopic Sample and SDSS Imaging Sample, and
extend this type of analysis 
beyond luminous red galaxies (LRGs) to a wide range of galaxies in both stellar mass and color. 

The primary uncertainty in turning a deprojected three-dimensional cross-correlation function
at small scales into a merger rate is in estimating the mean time for two galaxies to merge as a
function of stellar mass and separation. There are different estimates of merger timescale:
free-fall time, orbital time, and dynamical friction time
(\eg, \citealt{Binney87a, Boylan-Kolchin07a, conroy07a, kitzbichler08a, Bundy09a, Loz10a}).
In this paper, we will use an approximation to the two kinds of merger times under the assumption of \cite{kitzbichler08a} 
and \cite{Binney87a} as our
standard estimate. 
Both of these times are likely to be an under-estimate of the mean
merger time, because some close pairs will not merge at all. Any under-estimate of the merger 
time leads to an over-estimate of the growth rate.

Throughout this paper, all magnitudes are AB, all apparent magnitudes are model Mag, all masses are stellar masses (in units of $\hMsun$),
all velocities are in units of $\km \, \s^{-1}$,
all radii of galaxies are $r_{90}$ which contain $90$ percent of the Petrosian flux \citep{Blanton03b, blanton09a},
and all volumes and distances are comoving, calculated for a cosmological world
model with $(\Omega_m,\Omega_{\lambda}) = (0.3, 0.7)$ and Hubble constant 
$H_0 \equiv 70 \, h_{70}\, \km \, \s^{-1} \, \Mpc^{-1}$.


\section{DATA}
\label{data}

We use the NYU Value-Added Galaxy Catalog (VAGC) V7.2 data \citep{blanton05a}, 
which is built from SDSS data, as our source of spectroscopic data. We use all the SDSS imaging data 
as our source for cross-correlation samples.
The SDSS is a survey of about $10^4$ square degrees
\citep{fukugita96a, gunn98a, gunn05a, York99a, York01a, stoughton02a, abazajian03a, abazajian04a}.


   \subsection{Spectroscopic Subsample} 

\label{Spectroscopic Subsample}

\begin{figure}
   \centering
   \includegraphics[width=1.\textwidth]{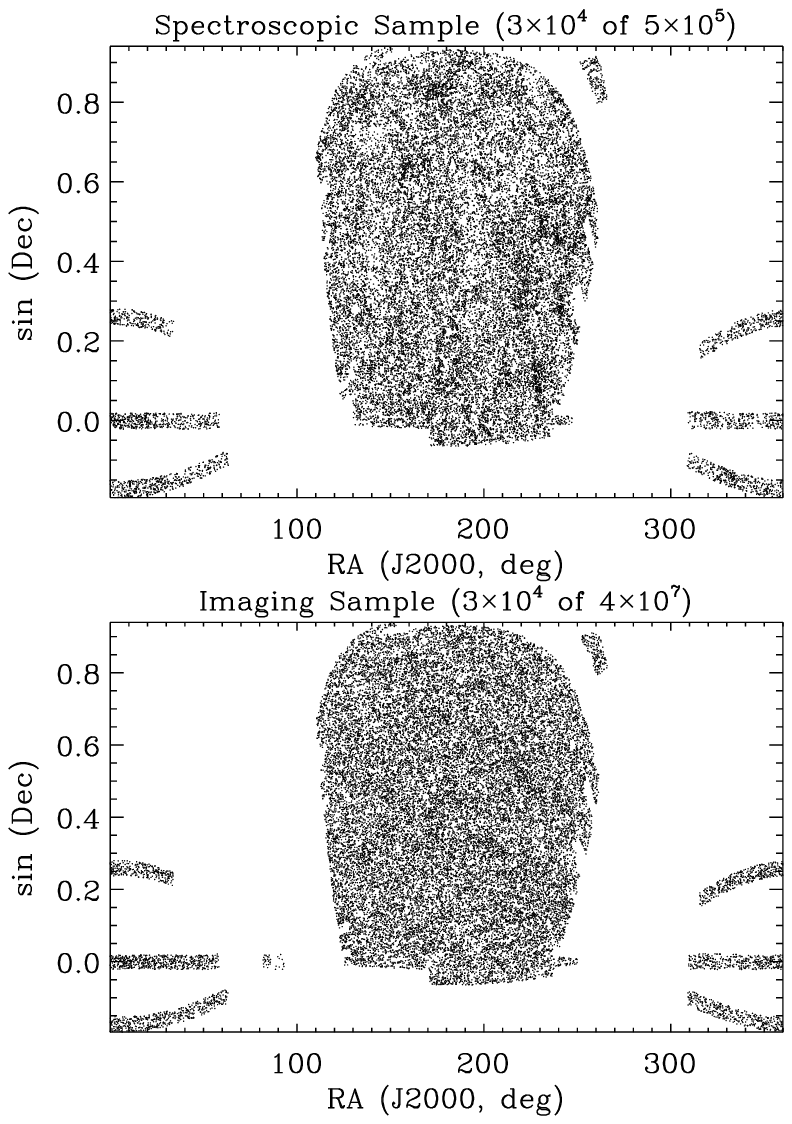}
   \caption{Sky distribution of the spectroscopic and imaging samples. For clarity, only a randomly chosen subsample of $3 \times 10^4 $ points is shown in each case.}
   \label{fig:decVSra}
   \end{figure}

\begin{figure}
   \centering
   \includegraphics[width=1.\textwidth]{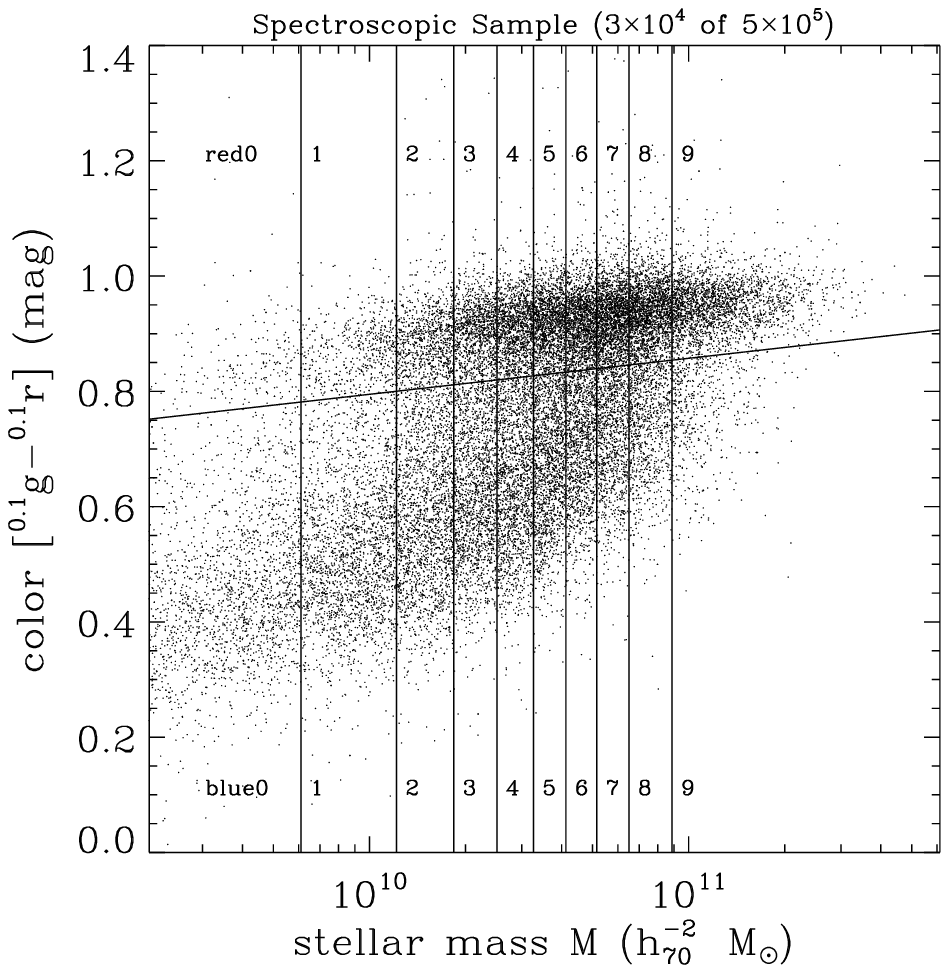}
   \caption{Distribution of the spectroscopic subsample in color and $K$-correct-estimated stellar mass. The nearly horizontal line separates galaxies into red and blue; the vertical lines separate galaxies into 10 subsamples with different stellar mass. For clarity, only a randomly chosen subsample of $3 \times 10^4 $ points is shown.}
   \label{fig:colorVSmass}
   \end{figure}

Our spectroscopic sample is drawn from NYU VAGC V7.2 data. We removed galaxies 
with apparent magnitudes $r < 14 \, \mag$, because SDSS photometric catalog missed 
many luminous galaxies nearby \citep{zhu10a}. Our spectroscopic sample contains about $8.6 \times 10^5$ main galaxies, 
of which about $5 \times 10^5$ SDSS Main Samples galaxies \citep{Strauss02a} are in the redshift 
range $0.03 < z < 0.15$ (see Figure~\ref{fig:decVSra}) with apparent magnitudes $r < 17.77$. 
We cut the redshift at $z < 0.15$,
because we want to avoid all quasars, and we cut the redshift at $z > 0.03$, 
because we want to avoid all stars.
We use the code {\tt sdss\_kcorrect} \citep{Blanton07a} to calculate 
the $K$-corrected absolute magnitude at $z = 0.1$ and stellar mass for each.
We cut this sample into 10 subsamples by stellar mass and make sure that in each subsamples
the number of galaxies is the same.
After that, we cut them into red and blue using the cut \citep{hogg04a}: 
\begin{equation} \label{eq:colorcut}
^{0.1}g - ^{0.1}r = 0.0625 \, \times \, \log \, (M) \, + \, 0.15
\end{equation}
where $M$ is the stellar mass.
These cut the spectroscopic sample into 20 spectroscopic subsamples; we call these subsamples $\setDDs$
(see Figure~\ref{fig:colorVSmass}).
For the red spectroscopic subsamples, we name the smallest stellar-mass subsample `red0', then name the second smallest stellar-mass subsample `red1',
and so on, so the largest stellar-mass subsample is named `red9', and the same for the blue spectroscopic subsamples.
Figures~\ref{fig:densityred} and \ref{fig:densityblue} show the number density of galaxies of the 18 spectroscopic subsamples
as a function of redshift.
The red0 and blue0 subsamples are low in number density, and highly affected by survey selection effects, 
so we do not use them further.
We can see that blue8, blue9 and red9 appear to rise in number density with redshift, 
this is because we removed galaxies with apparent magnitudes $r < 14 \, \mag$.

The SDSS suffers from the `fiber collision':
the angular separation between any two spectroscopic targets
must be larger than 55 arcsec.
There are about $\sim 9$ percent 
of target galaxies which do not have redshifts because of this fiber collision.
Using the counting-close-pairs technique of \cite{Masjedi08a},
the fiber collision will not affect our pair counts directly under 
the approximation that the unmeasured galaxies and the measured galaxies
are similar in cross-correlation with fainter galaxies.

Following the technique of \cite{Masjedi08a}, 
we estimate the weight $p_j$ that accounts the spectroscopic incompleteness from fiber collisions effects, and
the weight
$f_j $ that accounts the spectroscopic incompleteness from all other selection effects in SDSS.
For each galaxy $j$ in the spectroscopic subsample, we calculate $p_j$ by using a two-dimensional `FOF' (friends-of-friends) grouping algorithm
on our main galaxies targets with a 55-arcsec linking length:
\begin{equation} \label{eq:pj}
p_j = \frac {N_s^{j}} {N_{total}^{j}}
\end{equation}
Where $N_{total}^{j}$ is the total number of main galaxies targets in group $j$, 
$N_s^{j}$ is the number of main galaxies with redshift measurement in group $j$, and 
the group which contains galaxy $j$, is called group $j$.
All $p_j \ge 1$ because $N_{total}^{j} \ge N_s^{j}$.

For each spectroscopic subsample $\setDDs$, we create a random spectroscopic subsample $\setRRs$ which is 10 
times large as $\setDDs$.
For each galaxy in $\setDDs$, we created 10 galaxies with the exactly same redshift but angular position taken from
the two-dimensional random sample. 
Thus, we create a large random spectroscopic subsample $\setRRs$ which match the redshift distribution of $\setDDs$,
and $\setRRs$ is isotropic within the SDSS survey region.


The SDSS survey region is separated into small unique region `sector'.
For each random point $j$ in $\setRRs$:
\begin{equation} \label{eq:pj}
f_j = \frac {1} {F^{j}}
\end{equation}
Where
$F_j$
is the fraction of main galaxy targets for which a classification was obtained in the object's sector
(in our NYU-VAGC spectroscopic sample, the average $F_j$ is $\bar{F} \thickapprox 0.91 $).

Our correction of
fiber collisions is:
we weight the target $j$ in spectroscopic subsample $\setDDs$ as $p_j$
and weight the target $j$ in random spectroscopic subsample $\setRRs$ as $f_j$.
From the previous work \citep{Masjedi06a}, we know that this correction will improve in 
the spectroscopic incompleteness due to fiber collisions at very small
separations ($w(r_p) < 100 \hKpc$). 
In Section \ref{compare2zehavi}, 
we will compare our result with the result of \cite{Zehavi10a}, and will show that 
our result with this correction of
fiber collisions fits better  
than the result without this correction of fiber collisions.

For each spectroscopic sample `s' ($\setDDs$ or $\setRRs$), in which
there are $N_s$ spectroscopic galaxies,
we divide it into the 50 bins by lines of constant dec, so that
there are $N_s/50$ spectroscopic galaxies in each bin. 
Then we resample them into 50 leave-one-out resampling samples, so that 
there are $49/50 \times N_s$ spectroscopic galaxies in each sample.
We call them the `50 resampling samples', 
with which we can calculate our jackknife
resampling covariance matrix.


   \subsection{Imaging Subsamples}

\begin{figure}
   \centering
   \includegraphics[width=1.\textwidth]{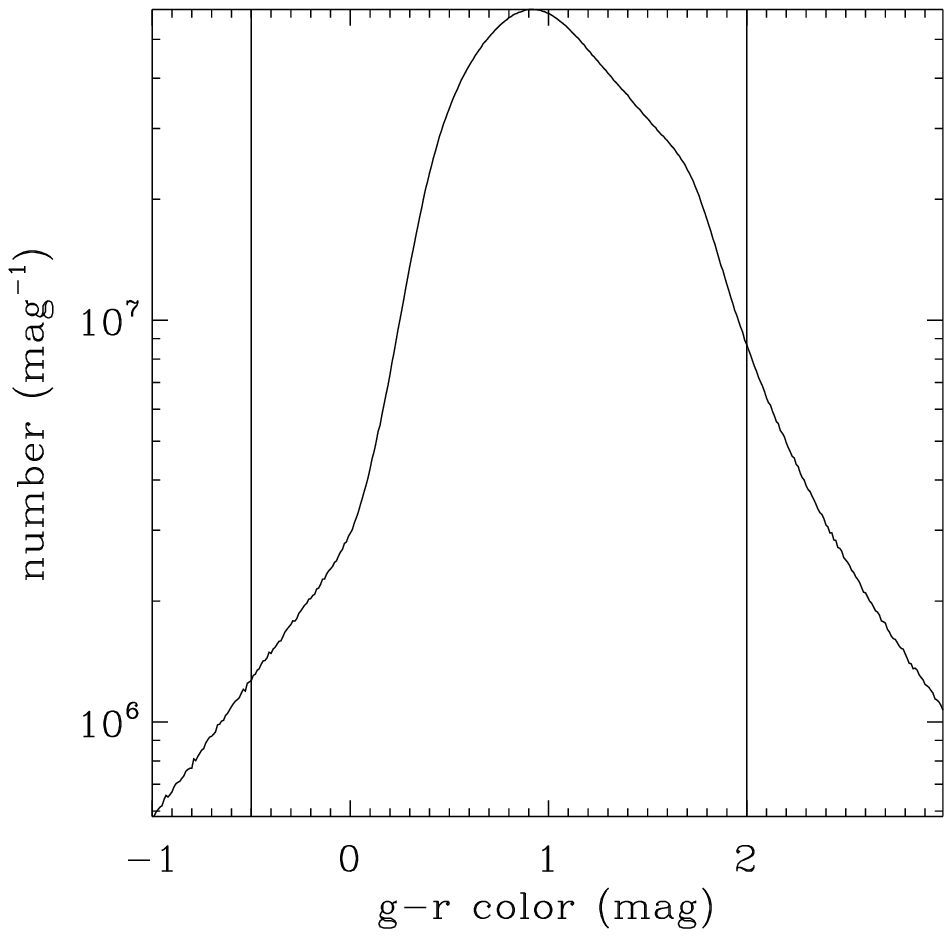}
   \caption{Distribution of $[g-r]$ color for the imaging sample. We use only galaxies from the imaging sample with  $-0.5 < [g-r] < 2.0 \, \mag$. }
   \label{fig:colorcut}
   \end{figure}

For our imaging data, we use a sample drawn from the full SDSS imaging catalog in which there are about
$4 \times 10^7$ galaxies. We include from the SDSS imaging sample only galaxies with
apparent magnitude $14 < r < 21.5 \, \mag$, and 
apparent color $ -0.5 < [g-r] < 2 \, \mag$, see Figures~\ref{fig:rcut} and \ref{fig:colorcut}.
We removed galaxies with apparent magnitudes $r < 14 \, \mag$ for the same reason with our spectroscopic sample.
We also removed galaxies with apparent magnitudes $r > 21.5 \, \mag$, because these galaxies are not  
well observed and their observed number density is much lower than their real number density.
Please note that this cut will affect our minor merger near mass ratio $1:100$.
Similar to the spectroscopic sample, 
we create a random imaging sample $\setRRi$ as large as possible.
The angular positions of galaxies in $\setRRi$ are taken from
the two-dimensional random sample. 
So $\setRRi$ is isotropic within the SDSS survey region.


   \subsection{Grid Method $K$-corrections for galaxies from the imaging sample}

We also need to compute the stellar mass of galaxies in the imaging subsamples in order to
determine the mass ratio, so we want to $K$-correct galaxies in the imaging subsamples.
However, we cannot $K$-correct individual galaxies in the imaging subsamples once and for all,
because we do not have spectroscopic redshifts for them. Each time we consider a pair
of galaxies, one from the spectroscopic subsample and one from the imaging subsample, we
assign the spectroscopic redshift to the galaxy from the imaging sample. This allows us to calculate
for each galaxy from the imaging sample in each spectroscopic--imaging pair a temporary $K$-corrected stellar mass 
and [$^{0.1}g \, - \, ^{0.1}r$] color for the purposes of that pair. We discard
these values and compute new ones when the galaxy from the imaging sample is used in another pair with
another galaxy from the spectroscopic sample.

\begin{figure}
   \centering
   \includegraphics[width=1.\textwidth]{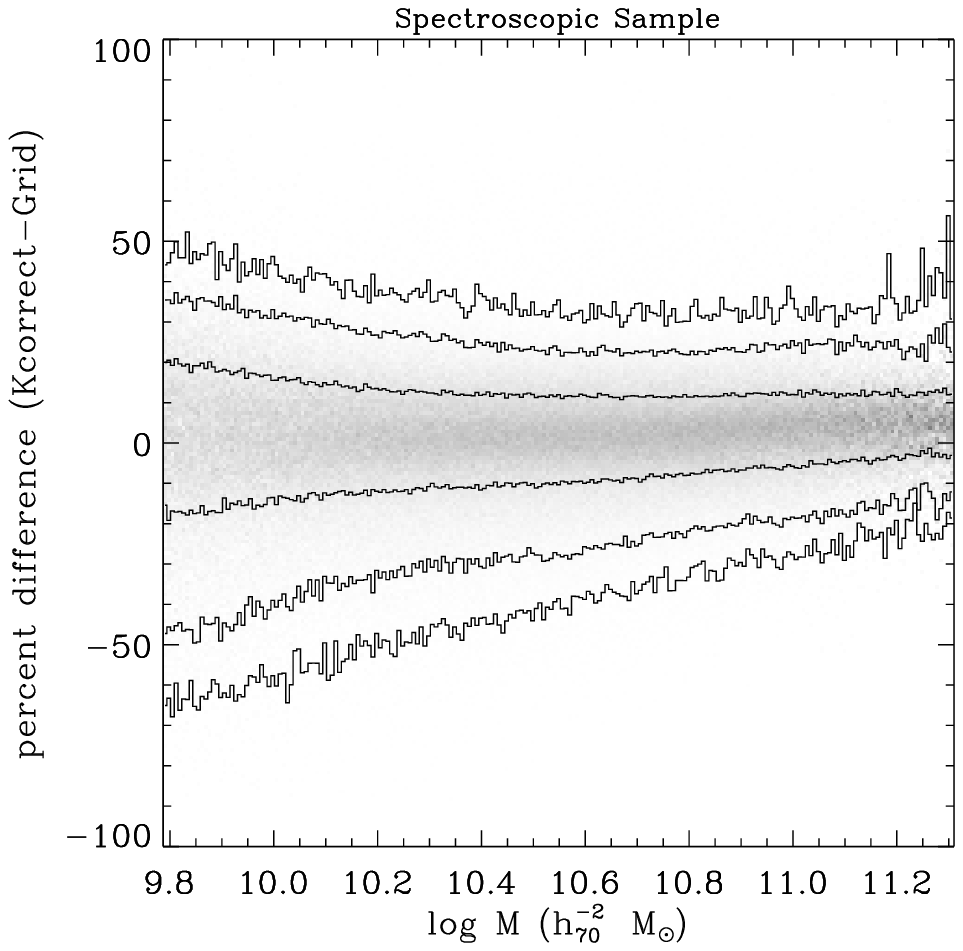}
   \caption{The difference between the $K$-correct-estimated stellar mass and the stellar mass estimated by the Grid Method. The contours show the 68, 95 and 99-percent intervals. The greyscale reflects the number of galaxies in each bin.}
   \label{fig:stellarmass}
   \end{figure}

To save time, we take galaxies from the NYU VAGC spectroscopic sample as representative
of all galaxy types, and apply the code {\tt sdss\_kcorrect} on the galaxies from a grid named `B' 
of observed $r$-band magnitude 
(0.5 mag per bin), [$g - r$] color (0.1 mag per bin), and redshifts between 0.03 and
0.15 (0.0002 per bin). We saved the mean $K$-corrected stellar mass $M_B$ and [$^{0.1}g_B \, - \, ^{0.1}r_B$] color
in a grid of observed $r$-band magnitude, [$g - r$] color, and redshift, also save the mean redshift $z_B$, mean $r_B$ and mean [$g_B - r_B$] color,
Thereafter we estimated
the $K$-corrected stellar mass and [$^{0.1}g \, - \, ^{0.1}r$] color for a 
galaxy G in grid B, and the [$g_G - r_G$] color of galaxy G is between [$g_B - r_B$] color of grid B and [$g_C - r_C$] color of grid
C which is next to grid B (that means $z_B$ and $z_C$ are the same, and $r_B$ and $r_C$ are the same):
\begin{equation} \label{eq:mass}
\log \, {M_G} = \log \, {(M_B \, \times \, \frac{(d_L^G)^2}{(d_L^B)^2}) } \, - \, \frac{r_G - r_B}{2.5} \, + \, \frac{ (\log \, {M_C} \, - \, \log \,{M_B}) \times ([g_G - r_G] - [g_B - r_B])} {[g_C - r_C] - [g_B - r_B]}
\end{equation}
\begin{equation} \label{eq:color}
[^{0.1}g_G \, - \, ^{0.1}r_G] = [^{0.1}g_B \, - \, ^{0.1}r_B] + [g_G - r_G] - [g_B - r_B]
\end{equation}
where $d_L$ is the luminosity distance calculated from redshift, and galaxy G is in grid B,
so the difference between $z_B$ and $z_G$ is small, similarly, the difference is small between 
$d_L^G$ and $d_L^B$, $r_b$ and $r_G$, [$g_B - r_B$] and [$g_G - r_G$].
because of these small differences, the equations (\ref{eq:mass}) and (\ref{eq:color}) can be used.
This speeds up the $K$-correction procedure immensely and only introduces a 12-percent one-sigma error for each
galaxy and there is little bias (see Figure~\ref{fig:stellarmass}), 
so only introduces percent-level errors in the results. 
We call this `Grid Method' hereafter.

For grids with observed $r$-band magnitude $r > 17.77 \, \mag$, we cannot get the mean stellar mass and
[$^{0.1}g \, - \, ^{0.1}r$] color directly, because there are
no galaxies from the spectroscopic sample at observed $r$-band magnitude $r > 17.77 \, \mag$.
In order to estimate the color and stellar mass of the galaxies in grid A with mean observed $r$-band magnitude
$r_A > 17.77 \, \mag$, we find the grid point B that has the nearest mean observed $r$-band magnitude 
$r_B$ ($r_B > 17.77 \, \mag$) and has the same observed
[$g - r$] color and the same redshift, also there are at least 10 galaxies from the spectroscopic sample in grid B. Then we estimate
$M_G$ and $[^{0.1}g_G \, - \, ^{0.1}r_G]$ color of galaxy G in grid A using the equation (\ref{eq:mass}) and (\ref{eq:color}).
If we cannot find a grid B satisfying the conditions, we will leave all the galaxies in grid A empty.

\begin{figure}
   \centering
   \includegraphics[width=1.\textwidth]{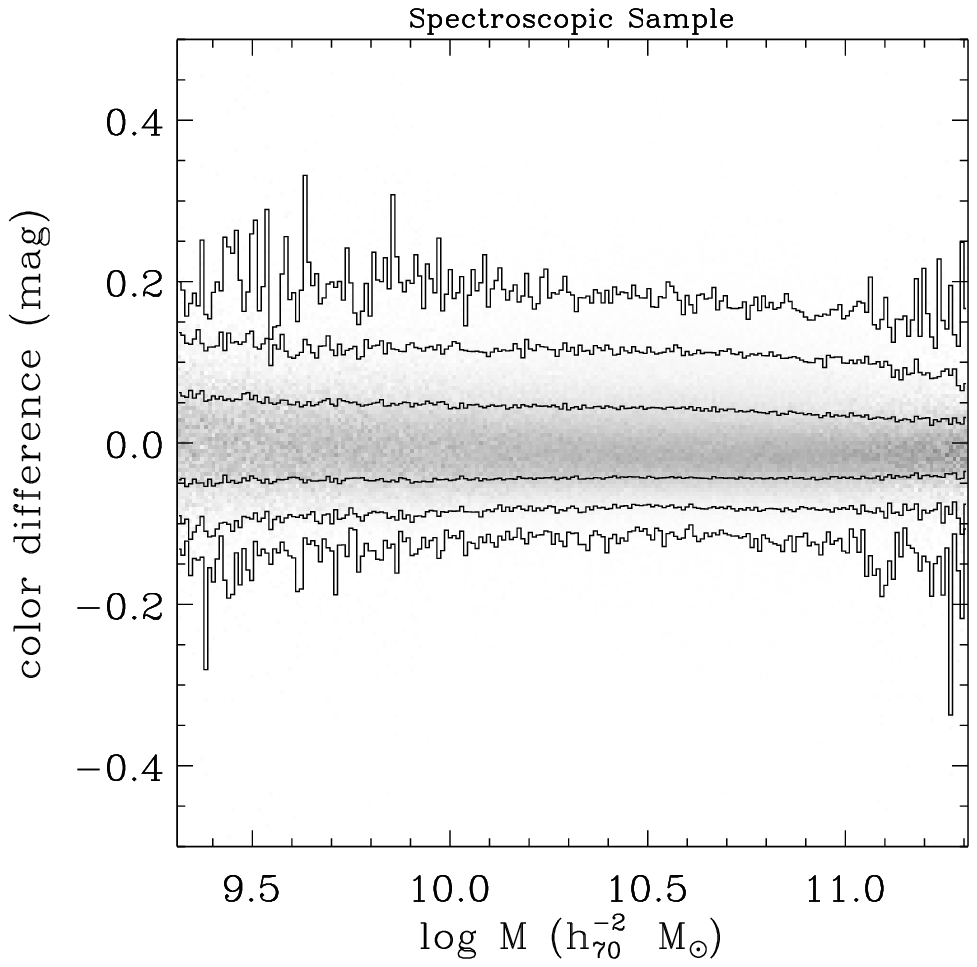}
   \caption{The difference between the $K$-correct-estimated $[^{0.1}g \,- \, ^{0.1}r]$ color and the $[^{0.1}g \,- \, ^{0.1}r]$ color estimated by the Grid Method. The greyscale and the contours are similar to those in Figure~\ref{fig:stellarmass}.}
   \label{fig:color}
   \end{figure}

Figure~\ref{fig:color} shows the difference between the color estimate using the Grid Method above 
and the color calculated using the code {\tt sdss\_kcorrect}. The one-sigma error for each
galaxy is about $0.04 \, \unit{mag}$ and there is little bias (see Figure~\ref{fig:color}).


   \subsection{Velocities for merger rate estimates}


In this Section, we will estimate the average orbital velocity which will be used to 
estimate the merger time in Section \ref{mergerrate}.
Please note that this approximation will induce a large error in the estimate of merger time, because we assume that
all close pairs merge under the following orbital velocity which is not true for close pairs in high velocity-dispersion.

\begin{figure}
   \centering
   \includegraphics[width=1.\textwidth]{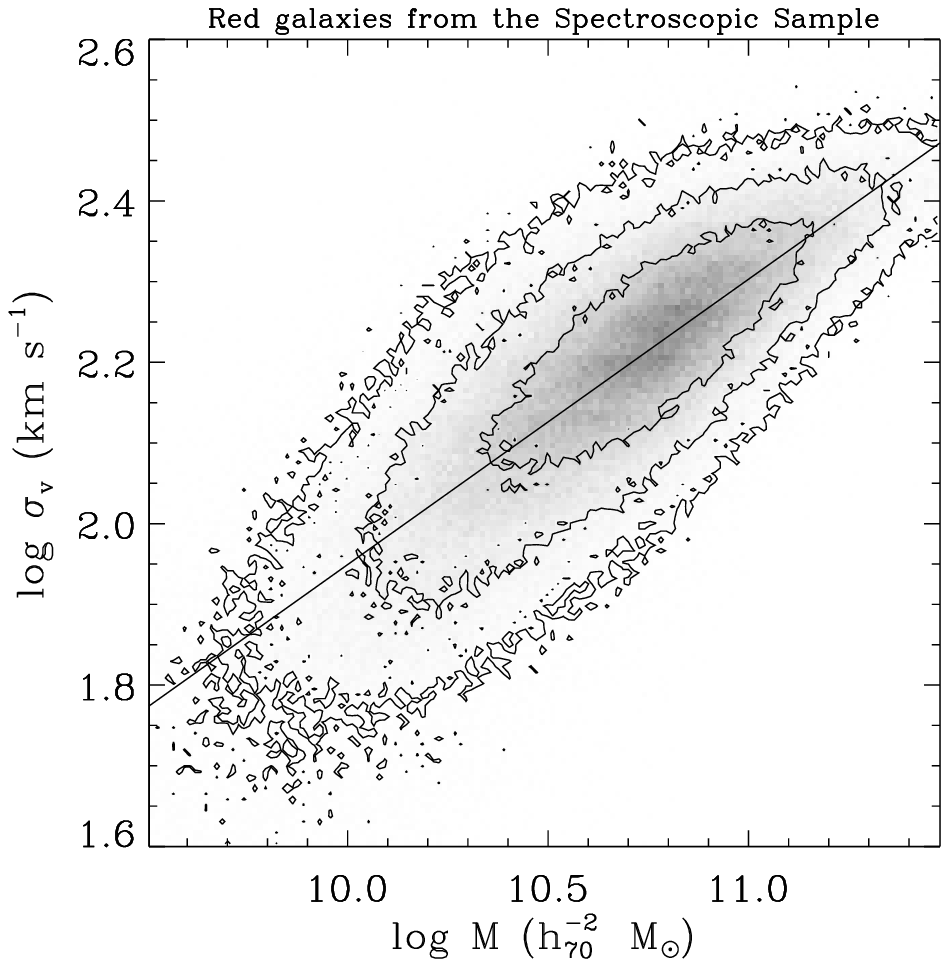}
   \caption{The relationship between $\sigma_{v}$, the stellar velocity dispersion and $M$, the stellar mass for red galaxies in the spectroscopic sample. The greyscale and the contours reflect the number of galaxies in each bin. The solid line is the linear fit, equation (\ref{eq:velocity_red}).}
   \label{fig:velocity}
   \end{figure}

The average orbital velocity for a galaxy from the imaging sample around a more massive
red galaxy from the spectroscopic subsample $s$ with average velocity dispersion $\sigma_v$ is very roughly $1.5$ times the
velocity dispersion, here we have included the factor of 1.5 to be conservative \citep{Masjedi08a}. 
We estimate the $\sigma_{v}$ with 
Faber-Jackson relation:
\begin{equation} \label{eq:velocity_red}
\log \, \sigma_{v} \, = \, a_1 \, + \, b_1 \, \log \, M_s^{red}
\end{equation}
where $M_s^{red}$ is the mean stellar mass of the red galaxies from 
spectroscopic subsample $s$. We performed a linear fit to the data
to obtain $a_1 = -1.588$ and $b_1 = 0.354$ (Figure~\ref{fig:velocity}).
Please note that the method above will induce a small enough error ($<1 \, \unit{percent}$) into our final result.

We estimate the average orbital velocity $V_{c}$ for a galaxy from the imaging sample around a more massive
blue galaxy from the spectroscopic subsample $s$ with the
Tully-Fisher relation:
\begin{equation} \label{eq:velocity_blue}
\log \, V_{c} \, = \, a_2 \, + \, b_2 \, \log \, \left<L_I\right>
\end{equation}
where $\left<L_I\right>$ is the mean I-band luminosity of the blue galaxies from 
spectroscopic subsample $s$, calculated from $L_I = M/r_I$, where $M$ is the stellar mass of the galaxy and 
$r_I$ is the I-band mass-to-light ratio of the galaxy calculated from the code {\tt sdss\_kcorrect}. 
We used for this relationship $a_2 = -0.835$ and $b_2 = 0.291$ \citep{courteau07a}.

Table \ref{tab:data0} provides this information for all 20 spectroscopic subsamples.

   
\section{Method}



For each spectroscopic subsample $\setDDs$, we cut the imaging sample into 16 subsamples by stellar mass: 
$10^{-\frac{j}{8}} < M_i^{G}/M_s < 10^{-\frac{j-1}{8}}$, where $M_i^{G}$ is the stellar mass 
of a galaxy from the imaging sample, calculated by the Grid Method using the redshift of the
spectroscopic galaxy, and $M_s$ is the mean stellar mass of the galaxies from 
spectroscopic subsample $s$, and $j$ is an integer $1 \le j \le 16$,
that means a mass ratio of $1:1$ to $1:100$ is covered.
After that, we cut them into red and blue using equation (\ref{eq:colorcut}) by
$[^{0.1}g \, - \, ^{0.1}r]$ color calculated by Grid Method.
For each spectroscopic subsample $\setDDs$, these cut the imaging sample into 32 imaging subsamples $\setDDi$.

In this section, we will show our method to estimate the merger rate between galaxies in spectroscopic subsample $\setDDs$
and galaxies in imaging subsample $\setDDi$:
\textsl{(1)}~we estimate the projected two-dimensional cross-correlation function $w_{si}(r_p)$ as a function of tangential projected separation $r_p$;
\textsl{(2)}~we de-project the smooth fit for the cross-correlation function $w_{si}(r_p)$ to obtain the three-dimensional
real-space cross-correlation function $\xi_{si}(r)$ as a function of real-space separation $r$;
\textsl{(3)}~we
estimate the merger rate using $\xi_{si}(r)$ and our
two kinds of merger times \citep{Binney87a, kitzbichler08a}.
In addition, we will also discuss our method of photometry correction.


   \subsection{Projected cross-correlation function}

\label{wp}


To estimate the $w_{si}(r_p)$ between
spectroscopic subsample $\setDDs$
and imaging subsample $\setDDi$,
we can integrate $\xi_{si}(r)$ along the line of sight (\eg, \citealt{davis83a}):
\begin{equation} \label{eq:wp}
w_{si}(r_p) = 2\int_0^{\infty} \di {y} \, \xi_{si} \, [ (r_{p}^2 \, + \, y^2) ^{1/2} ]
\end{equation}
This integral is dominated by scales $y \lesssim r_p$.


Using the previous approach \citep{Masjedi06a, Masjedi08a}, we estimate not $w_{si}(r_p)$ but 
$\rho_i \, w_{si}(r_p)$, where $\rho_i$ is the average comoving three-dimensional stellar mass density of the imaging 
subsample $i$:
\begin{equation} \label{eq:mwp}
\rho_{i} \, w_{si}(r_p) = \frac{D_{s}D_{i}}{D_{s}R_{i}} - \frac{R_{s}D_{i}}{R_{s}R_{i}}
\end{equation}
Where $D_{s}$ and $D_{i}$
represent the spectroscopic and imaging data subsamples, and $R_{s}$ and $R_{i}$
represent the spectroscopic and imaging random subsamples.
Similar to the previous method \citep{Masjedi06a, Masjedi08a}, 
equation (\ref{eq:mwp}) measures the mass-weighted abundance of pairs ($\frac{D_{s}D_{i}}{D_{s}R_{i}} $)
and subtracts the mean background level ($\frac{R_{s}D_{i}}{R_{s}R_{i}}$).
This method has been well tested in \cite{Masjedi06a}.

In detail, the factors are defined as follows:
\begin{equation} \label{eq:dd}
D_{s}D_{i} = \frac{\displaystyle\sum_{j \in \setDDs\setDDi}p_j}%
              {\displaystyle\sum_{j \in \setDDs}p_j} \quad 
\end{equation}
where the top sum counts the weighted pairs $j$ of galaxies from $\setDDs$ and $\setDDi$ separated by
tangential projected distance $r_p$, and the bottom sum is over
galaxies $j$ from $\setDDs$.
This factor $D_{s}D_{i}$ is dimensionless.
\begin{equation} \label{eq:drm}
D_{s}R_{i} = \frac{\displaystyle\sum_{j \in \setDDs\setRRi}p_j}%
              {\displaystyle\sum_{j \in \setDDs}p_j
\,\left[\frac{\di \Omega}{\di A}\right]_j\,\frac{\di M}{\di \Omega}} \quad
\end{equation}
where the top sum counts the weighted pairs $j$ of galaxies from $\setDDs$ and $\setRRi$ separated by
tangential projected distance $r_p$.
In the bottom, $\biggl(\frac{\di \Omega}{\di A} \biggr)_j$ 
is the inverse square of the transverse comoving distance \citep{hogg99a} to galaxy $j$ from $\setDDs$, and
$\frac{\di M}{\di \Omega}$ is calculated by $\frac{\di M}{\di \Omega} = \frac{\di N}{\di \Omega} \times M_i$,
where $\frac{\di N}{\di \Omega}$ is the two-dimensional number 
density of $\setRRi$ per solid angle, and
\begin{equation} \label{eq:mims}
M_i = M_s \times 10^{-\frac{j \, - \, 0.5}{8}}
\end{equation}
where $M_i$ is the mean stellar mass 
of galaxies from the imaging subsample $i$, and $j$ is an integer $1 \le j \le 16$.
Then $\biggl(\frac{\di \Omega}{\di A} \biggr)_j \times \frac{\di M}{\di \Omega}$ 
represents the average stellar mass of galaxies in $\setRRi$ 
per unit comoving area around each galaxy from $\setDDs$. 
This factor $D_{s}R_{i}$ has dimensions of comoving area divided by stellar mass.
\begin{equation} \label{eq:rd}
R_{s}D_{i} = \frac{\displaystyle\sum_{j \in \setRRs\setDDi}f_j}%
              {\displaystyle\sum_{j \in \setRRs}f_j} \quad 
\end{equation}
this is similar to equation (\ref{eq:dd}), but $\setRRs$ represents now the random catalog mentioned in Section
\ref{Spectroscopic Subsample}.
This factor $R_sD_i$ is dimensionless.
\begin{equation} \label{eq:rrm}
R_{s}R_{i} = \frac{\displaystyle\sum_{j \in \setRRs\setRRi}f_j}%
              {\displaystyle\sum_{j \in \setRRs}f_j
\,\left[\frac{\di \Omega}{\di A}\right]_j \frac{\di M}{\di \Omega}} \quad
\end{equation}
this is similar to equation (\ref{eq:drm}), but $\setRRs$ represents now the random catalog mentioned in Section 
\ref{Spectroscopic Subsample}. 
This factor $R_{s}R_{i}$ has dimensions of comoving area divided by stellar mass.

For some experiments, we need to estimate $n_{i} \, w_{si}(r_p)$, where $n_{i}$ is the average comoving three-dimensional number density of the imaging 
subsample $i$. We estimate this by the following estimation:
\begin{equation} \label{eq:nwp}
n_{i} \, w_{si}(r_p) = \frac{D_{s}D_{i}}{[D_{s}R_{i}]_N} - \frac{R_{s}D_{i}}{[R_{s}R_{i}]_N}
\end{equation}

For $D_sD_i$ and $R_sD_i$, it is as the same as equation (\ref{eq:dd}) and equation (\ref{eq:rd}). For $[D_sR_i]_N$ and $[R_sR_i]_N$:
\begin{equation} \label{eq:dr}
[D_{s}R_{i}]_N = \frac{\displaystyle\sum_{j \in \setDDs\setRRi}p_j}%
              {\displaystyle\sum_{j \in \setDDs}p_j
\,\left[\frac{\di \Omega}{\di A}\right]_j\,\frac{\di N}{\di \Omega}} \quad
\end{equation}
this is similar to equation (\ref{eq:drm}), but $\frac{\di N}{\di \Omega}$ is the two-dimensional number 
density of the random imaging catalog per solid angle. 
This factor $D_{s}R_{i}$ has dimensions of comoving area.
\begin{equation} \label{eq:rr}
[R_{s}R_{i}]_N = \frac{\displaystyle\sum_{j \in \setRRs\setRRi}f_j}%
              {\displaystyle\sum_{j \in \setRRs}f_j
\,\left[\frac{\di \Omega}{\di A}\right]_j \frac{\di N}{\di \Omega}} \quad
\end{equation}
this is similar to equation (\ref{eq:rrm}). 
This factor $R_sR_i$ has dimensions of comoving area.

In range of our interest $0.0149 <  r_p < 11.9 \, \hMpc$,
we bin the spectroscopic-imaging pairs counting by the comoving projected separation $r_p$ of the pair
where $r_p = r_k = 0.0149 \times 10^{k/5} \, \hMpc$ and $k$ is an integer $0 \le k \le 14$. 
We have already discussed how to bin the spectroscopic sample $\setDDs$ and the imaging sample $\setDDi$
in Section \ref{data}.
We have combined the 16 stellar mass bins into 4 to simplify the figures.
Figures~\ref{fig:nwpredWITHred} through \ref{fig:nwpblueWITHblue} show the 
results of our measurements of $\rho_{i} \, w_{si}(r_p)$.

The uncertainties on the results shown in these figures are estimated using jackknife resampling covariance matrix with 50 resampling samples (see Section
\ref{Spectroscopic Subsample}), please note that all the error bars in our graphs only 
come from the jackknife resampling covariance matrix (there are some other errors like the error of the color 
and stellar mass estimated by the Grid Method, and so on).
On hundreds of kiloparsec scales, the error bars for each subsample
are smallest.
On smaller scales, the error bars become larger because of the `shot
noise': the smaller the separations, the fewer the pair counts.
On larger scales ($> 1 \, \hMpc$), the error bars become larger because 
there are more and more
interlopers on larger scales which means that the background subtraction is more noisy.

Figures~\ref{fig:nwpredWITHred} through \ref{fig:nwpblueWITHblue}
show that $n_i \, w_{si}(r_p)$ is a complex function of $r_p$. 
However, on very small scales - tens of kiloparsec scales, we assume that $n_i \, w_{si}(r_p)$
scales (something) like $r_p^{-1}$ \citep{Masjedi06a, Masjedi08a}.
We fit each set of $w_{si}(r_k)$ data with the smooth model:
\begin{equation} \label{eq:wrp0}
\widetilde{w}(r_p) = w_{0} \left[1 \, + \, \frac{r_p}{r_{c}}\right]^{\gamma} \, \left[\frac{r_p}{r_{c}}\right]^{-1}
\end{equation}
by minimizing 
\begin{equation} \label{eq:chi2}
\chi_{si}^2 \equiv \sum_{k} \frac{(\rho_{i} \, \widetilde{w}(r_k) \, - \, \rho_{i} \, w_{si}(r_k))^2}{\rho_{i}^2 \, \sigma_{si}(r_k)^2}
\end{equation}
We 
choose $r_c = 12.5 \, r_{90}$ 
for Figure~\ref{fig:nwpredWITHred} to Figure~\ref{fig:nwpblueWITHblue} (where $r_{90}$ is the
median radii of the galaxies in the corresponding spectroscopic subsamples); 
then find the $\gamma$ and $\rho_i \, w_{0}$ that 
minimizes $\chi_{si}^2$. Figures~\ref{fig:nwpredWITHred} 
through \ref{fig:nwpblueWITHblue} show us these fits.


   \subsection{Photometry Correction}
\label{recover}

One important issue with all clustering measurements on small scales
is possible photometric biases when measuring close pairs. 
This issue can directly lead to biased flux measurements \citep{Masjedi06a}
and biased color measurements for galaxies, and will indirectly affect stellar masses, k-corrections, etc.
This
can be due to poor photometry in crowded systems \citep{Patton11a}.

\begin{figure}
   \centering
   \includegraphics[width=1.\textwidth]{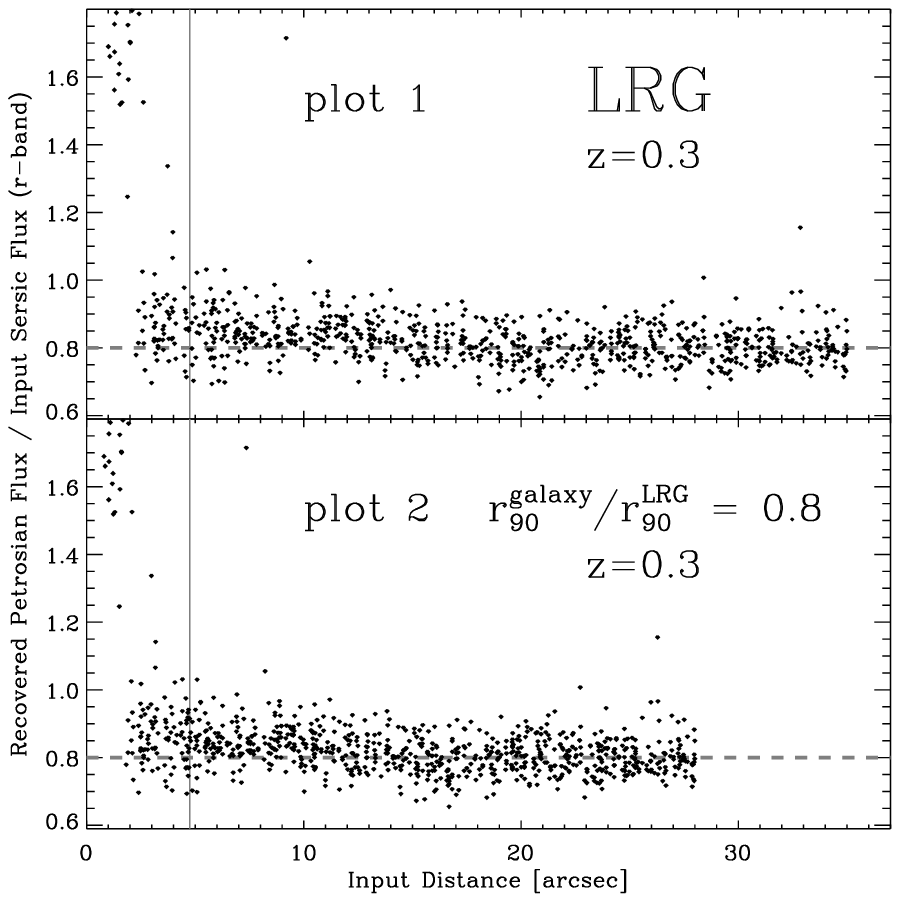}
   \caption{Recovered Petrosian flux to input S$\mathrm{\acute{e}}$rsic
flux as a function of the separation of the two galaxies in the pair. 
We show both LRG and small galaxy whose radius is only $80 \percent$ of that of LRG.
The vertical line shows the smallest separation in our research at $z=.15$ and $r_p = 14.9 \hKpc$. For LRG, on
average there is an excess in the recovered flux of galaxies separated
by less than $20~\arcs$. }
   \label{fig:recover}
   \end{figure}

We build our method to correct photometric biases upon the photometry test of \cite{Masjedi06a}.
In \cite{Masjedi06a}, they created fake
images of pairs of identical galaxies with separations ranging from $2$ to
$35~\arcs$. These galaxies represent
passively evolving LRG galaxies observed at a redshift of $z=0.3$ with
de Vaucouleurs profiles ($n=4$ S$\mathrm{\acute{e}}$rsic profiles).
Then they placed one such galaxy pair onto RUN $2662$ (which has a typical SDSS seeing of about $1~\arcs$)
of SDSS imaging. 
After inputting the known info into the mock galaxy images, they
processed these images as raw SDSS images using the standard SDSS pipeline, PHOTO, to
determine the effect of proximity of galaxies on their measured
properties (see plot $1$ in Figure~\ref{fig:recover}).
At separations larger than $20~\arcs$, the Petrosian flux
measures $79.5$ percent of the input S$\mathrm{\acute{e}}$rsic flux,
which is calculated by three sigma out-layer rejected average. In other words, the
Petrosian flux only measures about $80$ percent of a galaxy's
light. We are interested in intermediate separations, ($ 5 < s < 20~\arcs$), in which the
fraction of the recovered flux to input flux increases to $83$
percent.  This increase is likely due to a double counting of the low
level diffuse emission from the two galaxies which is being poorly
deblended between the two objects. 

\begin{figure}
   \centering
   \includegraphics[width=1.\textwidth]{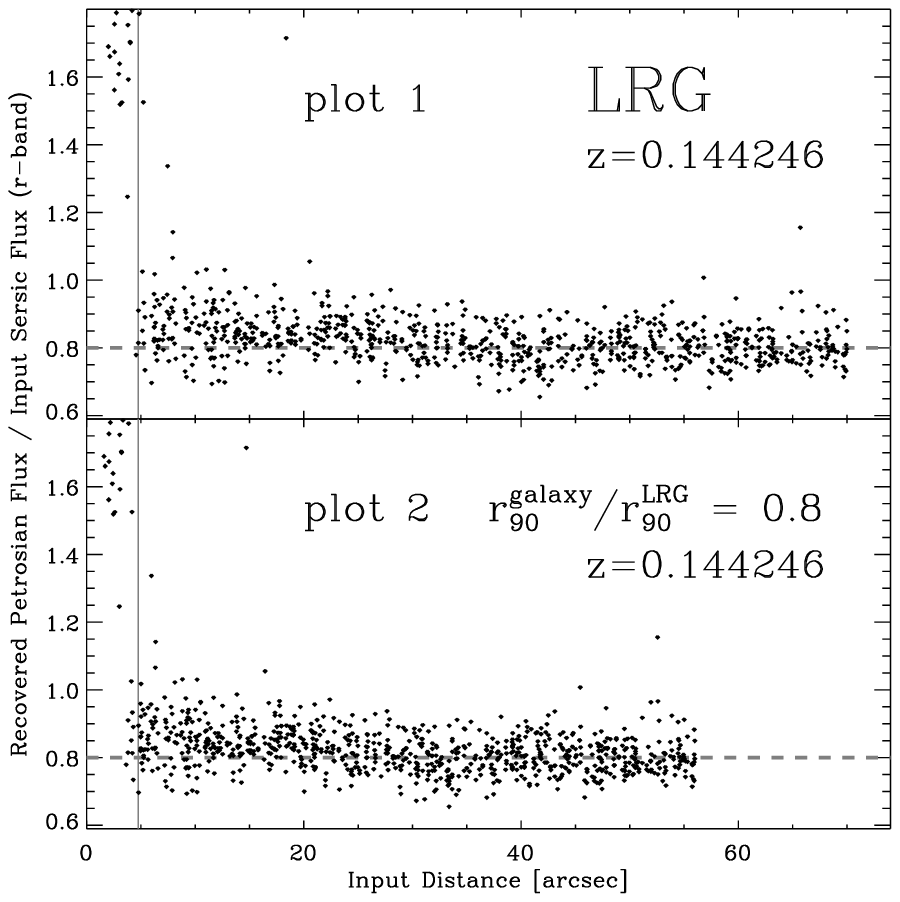}
   \caption{Recovered Petrosian flux similar as Figure~\ref{fig:zehavi_wp}.}
   \label{fig:recover2}
   \end{figure}

\begin{figure}
   \centering
   \includegraphics[width=1.\textwidth]{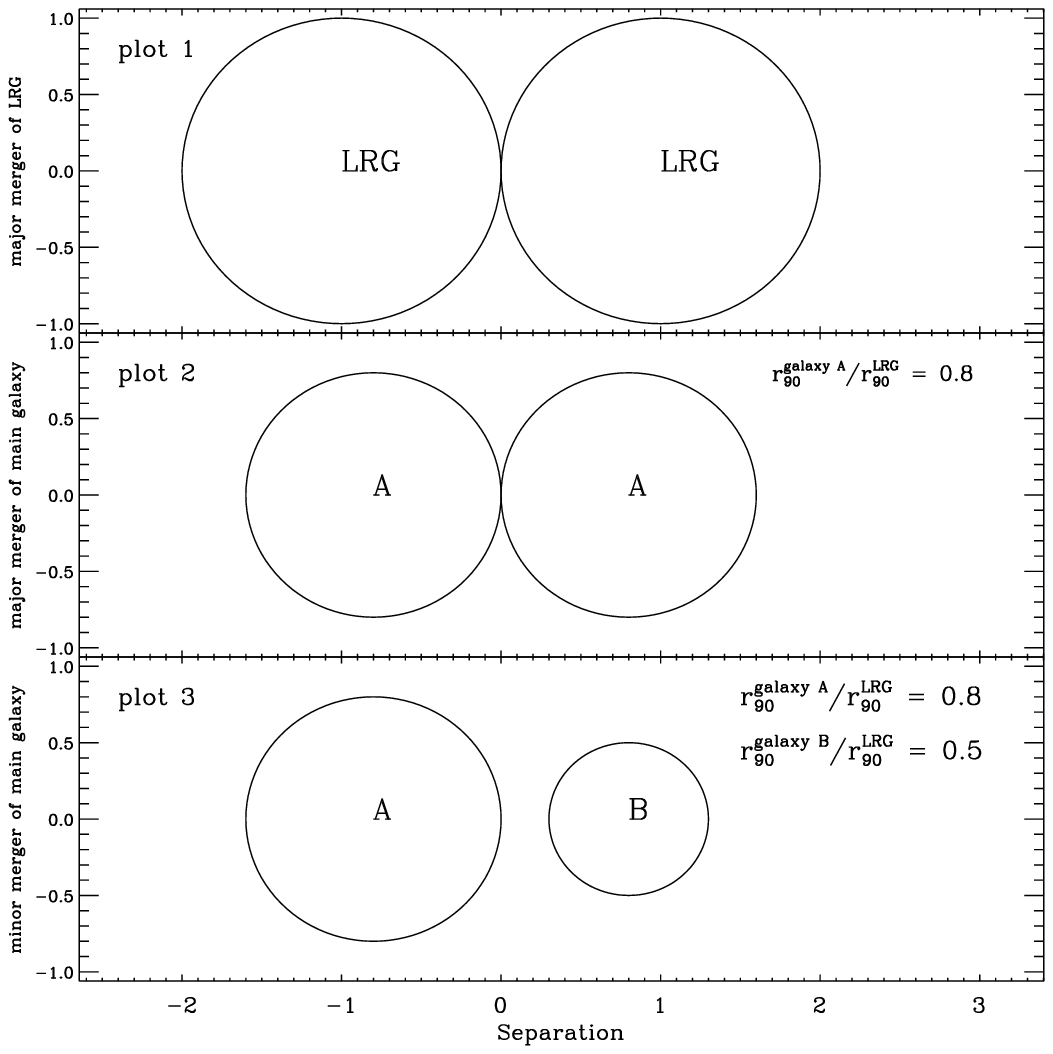}
   \caption{We show the major merger of LRG \citep{Masjedi06a}, major merger of main galaxy and minor merger of main galaxy. 
Please note that this is only a sketch, the radius and separation may be much different.}
   \label{fig:r90}
   \end{figure}

For pairs of main galaxies, we study two different cases: one for galaxy pairs
consisting of two identical galaxies and another with galaxies of
different stellar mass.
For Case 1, we consider a pair of identical main galaxies with radii of $r_{90}^{main}$ 
(at redshift $0.03< z_{main} < 0.15$)
than that of the LRGs $r_{90}^{LRG}$ at redshift $z = 0.3$ (see plot $1$ and $2$ in Figure~\ref{fig:r90}).
We take the following two approximations:
\textsl{(1)}~If we consider a pair of identical main galaxies which are at the same redshift $z=0.3$, then the
only difference between this pair of main galaxies and the pair of LRGs (see plot $1$ in Figure~\ref{fig:recover})
is that the angular radii of the main galaxies is smaller than that of LRGs
by a factor of $r_{90}^{main}/r_{90}^{LRG}$.
So we compress the result of plot 1 in Figure~\ref{fig:recover} by a factor of $r_{90}^{main}/r_{90}^{LRG}$ 
(see plot $2$ in Figure~\ref{fig:recover} as an example) as
the effect of proximity of a pair of identical main galaxies.
\textsl{(2)}~If we consider a pair of LRGs which are at a different redshift $z = z_{main}$, then the
only difference between this pair of LRGs and the pair of the LRGs at redshift $z=0.3$ (see plot $1$ in Figure~\ref{fig:recover})
is that the angular radii of this pair of LRGs is larger than that of the LRGs at redshift $z=0.3$
by a factor of $D^{main}/D^{LRG}$,
where $D^{main}$ is the comoving distance from $z = z_{main}$ to us, and $D^{LRG}$ is the comoving distance from $z = 0.3$ to us.
So we stretch the result of plot 1 in Figure~\ref{fig:recover} by a factor of $D^{main}/D^{LRG}$ 
(see plot $1$ in Figure~\ref{fig:recover2} as an example, and $D^{main}$ at redshift $z= 0.144246$ is nearly half of $D^{LRG}$) as
the effect of proximity of a pair of LRGs at redshift $z = z_{main}$.
Combining the above two approximations,
the final effect of proximity of a pair of identical main galaxies
at a different redshift
$z = z_{main} \ne 0.3$, will be stretched by a factor of $r_{90}^{main}/r_{90}^{LRG} \times D^{main}/D^{LRG}$
(see plot 2 Figure~\ref{fig:recover2} as an example).
Using this method we correct the flux measurement of our sample on
small scales of major merger between main galaxies and we assume that the correction of stellar mass is equal to that of flux.
For Case 2, it is based on the above Case $1$ but involving the radii $r_{90}$ of the pair of galaxies ($r_A$ and $r_B$) in our estimation.
In plot $2$ of Figure~\ref{fig:r90}, we take the third approximation:
\textsl{(3)}~The flux density from the left galaxy onto the right galaxy is a 
constant $D_A$ (this approximation lead to a small error comparing to the flux density from the galaxy A onto the galaxy B 
in plot $3$ of Figure~\ref{fig:r90}).
We mark the percent increase of flux from the other galaxy as $P_{flux}$, that of stellar mass as $P_{mass}$, radius of galaxy A as $r_A$ 
total flux as $F_A$ and stellar mass as $M_A$.
Then we get the following result: 
\begin{equation} \label{eq:recover}
P_{mass} = P_{flux} = \frac{D_A \times \pi \, r_A^2}{F_A}
\end{equation}
In plot $3$ of Figure~\ref{fig:r90}, we assume that the flux density from the galaxy A onto the galaxy B is equal to $D_A$ and
the flux density from galaxy B onto galaxy A is equal to $D_B$.
We mark the percent increase of flux of galaxy B from the galaxy A as $P_{flux}^B$, that of stellar mass as $P_{mass}^B$, radius of galaxy B as $r_B$ 
total flux as $F_B$, stellar mass as $M_B$, 
and the percent increase of flux of galaxy A from the galaxy B as $P_{flux}^A$, that of stellar mass as $P_{mass}^A$.
Then we get the following results: 
\begin{equation} \label{eq:recoverDB}
D_B = D_A \times \frac{F_B}{F_A} = D_A \times \frac{M_B}{M_A}
\end{equation}
\begin{equation} \label{eq:recoverA}
P_{mass}^A = P_{flux}^A = \frac{D_B \times \pi \, r_A^2}{F_A} = P_{mass} \times \frac{M_B}{M_A}
\end{equation}
\begin{equation} \label{eq:recoverB}
P_{mass}^B = P_{flux}^B = \frac{D_A \times \pi \, r_B^2}{F_B} = P_{mass} \times \frac{M_A}{M_B} \times \frac{r_B^2}{r_A^2}
\end{equation}
Using this result we correct the flux and stellar mass measurements of our sample on
small scales of minor merger between main galaxies.

Please note that our photometry correction is overestimated because of
the above approximations \textsl{(1)} and \textsl{(2)}.
In approximations \textsl{(1)} and \textsl{(2)}, we assume that the 
absolute angular scale does not matter; and the only change comes from the ratio of the 
absolute angular radii of the pair of galaxies to the absolute angular 
separation of the pair of galaxies.
However, we know that absolute angular scale \emph{does} matter: at the same ratio the larger
the absolute angular is, the easier the deblending will be.
So approximations \textsl{(1)} and \textsl{(2)} will contribute a few percent error 
in our final result of photometry correction.

After the above photometry correction, we reset our spectroscopic and imaging subsamples
to recalculate the $\rho_i \, w_{si}(r_p)$ using the method in
Section \ref{wp} (see the first and second data points in Figures~\ref{fig:nwpredWITHred} 
through \ref{fig:nwpblueWITHblue}). Please note that we also apply our photometry correction on the galaxies with no nearby companions,
however these galaxies have zero weight in our pair-counting and will not affect our result, because there are no 
companions near these galaxies during counting pairs. 
In order to show our method of photometry correction is robust, we double our photometry correction and find that all the percentage difference
between the result of our photometry correction and double our photometry correction is below $26$ percent for one data point.
Then, this data point with $26$ percent change will only contribute a few percent error
in our final result of 
the total fractional accretion rate after our fitting curve (see equation (\ref{eq:wrp0}) and (\ref{eq:chi2})).
So, if we assume that the percent error of our photometry correction in flux is $100 \percent$, the final effect
onto the total fractional accretion rate is at most a few percent.

We use the above method to correct the photometric biases, and we find that our 
correction due to photometric biases is much smaller than that of 
\cite{Masjedi06a}, because: 
\textsl{(1)}~Main galaxies have smaller radii than LRGs, so it is easier to deblend a pair of main galaxies than a pair of LRGs. 
\textsl{(2)}~Photometry correction of auto-correlation of \cite{Masjedi06a} is larger than that of 
our cross-correlation, this is because of the difference of the stellar mass cut of spectroscopic/imaging sample
between us: for auto-correlation, the stellar mass cut of spectroscopic/imaging sample will be $M_s>M_{threshold}^{lower}$
and $M_i>M_{threshold}^{lower}$, so after photometry correction, the only effect is that some spectroscopic (and imaging)
galaxies near $M_{threshold}$ will be cut off from the spectroscopic (and imaging) sample which will decrease
the pair-counting. However this is not the only effect on cross-correlation.
For cross-correlation, the stellar mass cut of spectroscopic and imaging sample will be $M_{threshold}^{lower}<M_s<M_{threshold}^{upper}$
and $M_{i}^{lower}<M_i<M_{i}^{upper}$, so after photometry correction, besides the above effect
there is another effect that some spectroscopic and imaging
galaxies which are a little bit above $M_{threshold}^{upper}$ or $M_{i}^{upper}$ will be 
counted into the spectroscopic/imaging sample from outside. This effect will increase
the pair-counting.
Combining the two effects above for cross-correlation, the final photometry correction for cross-correlation
will be smaller than that for auto-correlation.


   \subsection{Three-dimensional statistics}

The smooth fit $w(r_p)$ to each projected correlation functions $w_{si}(r_p)$ can be 
deprojected to get an estimate of the three-dimensional space correlation function $\xi(r)$ by
\begin{equation} \label{eq:nkxi}
\rho_{i} \, \xi(r) = -\frac {1}{\pi}\int_{r}^{\infty} \di r_p \frac{\di [\rho_{i} \, w_p(r_p)]}{\di r_p}(r_p^2-r^2)^{-1/2}
\end{equation}
\citep[\eg,][]{davis83a}, where $\rho_i$ is a constant.

The mean total stellar mass $M_i^*$ of galaxies from a specific imaging
subsample $i$ within a given small three-dimensional separation $r_{close}$ around each galaxy from
spectroscopic subsample $s$ is:
\begin{equation} \label{eq:ni_temp}
M_i^* = \rho_i\int \di V_i \, [1 \, + \, \xi_{si}(r)] = 4 \, {\pi} \, \rho_i\int_{0}^{r_{close}} r^2 \, \di r \, [1 \, + \, \xi_{si}(r)]
\end{equation}
At small scales, $\xi_{si}(r) \gg 1$, so:
\begin{equation} \label{eq:Ni}
M_i^* \thickapprox 4 \, {\pi} \, \int_{0}^{r_{close}} r^2 \, \di r \, [\rho_i \, \xi_{si}(r)] 
\end{equation}
From $[\rho_i \, \xi_{si}(r)] $ we can see that we do not need to measure $\rho_i$ and $w_{si}(r_p)$ separately.

At very small scale ($r_{close} \ll r_c$):
\begin{equation} \label{eq:Mi0}
M_i^* \thickapprox 4 \, \rho_{i} \, w_{0} \, r_c \, r_{close}
\end{equation}

Similarly, the average number $N_i^*$ of galaxies from a specific imaging
subsample $i$ within a given small three-dimensional separation $r_{close}$ per galaxy from
spectroscopic subsample $s$ at very small scale ($r_{close} \ll r_c$) is:
\begin{equation} \label{eq:nitemp}
N_i^* \thickapprox 4 \, n_{i} \, w_{0} \, r_c \, r_{close}
\end{equation}


   \subsection{Merger rate}
   
\label{mergerrate}

We can estimate the merger rate $\Gamma_i$ of galaxies from sample $i$ into galaxies from sample $s$ per 
spectroscopic galaxy per unit time by:
\begin{equation} \label{eq:merger}
\Gamma_i = \frac{N_i^*}{t_{merge,i}}
\end{equation}
The mean fractional stellar-mass accretion rate of galaxies from spectroscopic subsample $s$ from merging with
galaxies from imaging subsample $i$ per unit time is:
\begin{equation} \label{eq:growth}
\left[\frac{\di \, \mathrm {ln} M_s}{\di t}\right]_i = \frac{1}{M_s}\left[\frac{\di M_s}{\di t}\right]_i \thickapprox \frac{M_i^*}{t_{merge,i} \, M_s}
\end{equation}

In principle, all merger rate estimates depend on the radius
$r_{close}$ inside of which we have counted close pairs. 
However in this work, we are interested in the instant merger rate estimates, 
which means that we are interested in the range of $r_{close} \ll r_c$, and this
range can be reached using our fit lines.
Another reason why we use our fit lines instead of using the 
data points at very small scales, is that
the error bars due to the shot noise for each subsample are very large. 
From these fit lines in Figures~\ref{fig:nwpredWITHred} through
\ref{fig:nwpblueWITHblue} we know, 
over the range of interest ($r_{close} \ll r_c$)
$w_{si}(r_p)$ scales like $r_p^{-1}$,
$\xi(r)$ scales like $r^{-2}$, $N_i^*$ and $M_{i}^*$ scale (something)
like $r_{close}$. Similarly, both of the time-scales ($t_{KW,i}$ and $t_{BT,i}$)
scale like $r_{close}$. 
For this reason, at $r_{close} \ll r_c$ the above
merger and accretion rates 
do \emph{not} depend strongly on $r_{close}$.

In this work, we use these two merger time estimates: $t_{BT,i}$
from \cite{Binney87a} and $t_{KW,i}$ 
from \cite{kitzbichler08a}. 
Both of them depend on the orbital merger time $t_{orbit}$:
\begin{equation} \label{eq:orbit}
t_{orbit}^{red} = \frac{2 \, \pi \, r_{close}}{1.5 \, \sigma_v}
\end{equation}
Where $1.5 \, \sigma_v$ is the average orbital velocity for a galaxy from the imaging sample 
orbiting a more massive red galaxy from the spectroscopic sample with velocity
dispersion $\sigma_v$, see equation (\ref{eq:velocity_red}).
Similarly,
the orbital merger time for a galaxy from the imaging sample merged into a more massive
blue galaxy from the spectroscopic subsample is
\begin{equation} \label{eq:orbit_blue}
t_{orbit}^{blue} = \frac{2 \, \pi \, r_{close}}{V_c}
\end{equation}
Where $V_c$ is the average orbital velocity for a galaxy from the imaging sample around a more massive blue 
galaxy from the spectroscopic sample, see equation (\ref{eq:velocity_red}).

For the assumption of $t_{KW,i}$ from \cite{kitzbichler08a}, the approximation becomes
\begin{equation} \label{eq:dyntime_KW}
t_{KW,i} = t_{orbit} \left[\frac{M_s}{M_i}\right]^{0.3}
\end{equation}
where we assume $M_s > M_i$.
The solid lines in Figures~\ref{fig:growthred} and \ref{fig:growthblue} as a function of the 
mass ratio $M_i/M_s$ show the merger rate under assumption of \cite{kitzbichler08a}. 
The total fractional accretion rate is the area under each curve.

For the assumption of $t_{BT,i}$ from \cite{Binney87a}, $t_{BT,i}$ is longer than
the rbital merger time $t_{orbit}$ by a factor roughly equal to the ratio of the stellar masses \citep{Binney87a, Masjedi08a}:
\begin{equation} \label{eq:dyntime}
t_{BT,i} = t_{orbit} \frac{M_s}{M_i}
\end{equation}
where we assumed $M_s > M_i$.
The dashed lines in Figures~\ref{fig:growthred} and \ref{fig:growthblue} as a function of the 
mass ratio $M_i/M_s$ show the merger rate under assumption of all possible
mergers taking place within one dynamical friction time \citep{Binney87a, Masjedi08a}, 
and the total fractional accretion rate is the area under the curves.

We integrate the mass accretion rate from mergers over all imaging subsamples and find the total
fractional accretion rate ($\hhGyr$) of all the main galaxies. This is shown with solid lines (for
$t_{KW,i}$) and dashed lines (for $t_{BT,i}$) in Figure~\ref{fig:growth1}. Please note that the errors from 
merger time are shown in this figure.

Table \ref{tab:data1} shows the total fractional accretion rates of all the 20 spectroscopic subsamples for 
both the assumption of $t_{merge,i} = t_{KW,i}$ and the assumption of $t_{merge,i} = t_{BT,i}$.


\section{Comparison to Previous Work}

\label{compare}

   \subsection{Comparison to previous clustering results}

\label{compare2zehavi}

We estimate not $w_p(r_p)$ but $n_i \, w_p(r_p)$. In order to estimate $w_p(r_p)$ and compare 
with previous results, we need to estimate $n_i$:
\begin{equation} \label{eq:wpwithoutni}
w_p(r_p)=\frac{n_i \, w_p(r_p)}{n_i}
\end{equation}
It is difficult within the SDSS data to precisely
measure the real-space number densities for the imaging subsamples with stellar masses $M_i < 6 \times 10^9 \hMsun$,
because there is only good spectroscopic information about bright members of the imaging sample.
However, for galaxies with stellar masses $M_i > 6 \times 10^9 \, \hMsun$, the real-space number density $n_i$
for the imaging subsample $i$ is measurable:
\begin{equation} \label{eq:nitrue}
n_i=\frac{n_i^s}{\bar{F}}
\end{equation}
where $n_i^s$ is the average real-space number density for the corresponding spectroscopic subsample $s$ 
within its volume limit, and $\bar{F}$
is the mean fraction of Main targets for which a classification was obtained in the object's sector,
for NYU-VAGC spectroscopic sample $\bar{F} \thickapprox 0.91 $. We assume that $n_i$ is non-evolving 
over the redshift range of interest.
We find the lower and upper redshift limits of the volume-limit for the 
corresponding spectroscopic subsample $s$ in order to calculate $n_i^s$.

Galaxy clustering has been measured at intermediate and small scales 
\citep{Zehavi05a, Zehavi10a, Masjedi06a, Chen09a, Li10a, White11a}.
Our results are consistent with the results of \cite{Zehavi10a} (see Figure~\ref{fig:zehavi_wp}).
In order to generate our mass-threshold samples which are nearly the same as their luminosity-threshold samples,
we calculated $M^*$, the mean mass of the galaxies nearby their 
luminosity-threshold $M_r$ as our mass-threshold, and 
we cut $M_s > M^*$ and $M_i > M^*$ to generate the corresponding mass-threshold samples.
This turns our cross-correlation into an auto-correlation. In order to calculate $n_i^s$ for
the two subsamples with $M_r < -18.0$ and $M_r < -18.5$ in Figure~\ref{fig:zehavi_wp}, 
we use the peak real-space 
number densities instead of the average real-space number densities, because there
are no obvious volume-limit in these two subsamples.

Similarly, we can measure $\rho_{i}$ and $w_{si}(r_p)$ separately 
instead of $\rho_{i} \, w_{si}(r_p)$.

In in order to show that our result successfully corrected the fiber collisions,
we compare our result with the extension of the best-fit power
law from \cite{Zehavi10a}, 
see Figure~\ref{fig:zehavi_wp2} and \ref{fig:zehavi_wp3}. 
The extension dashed line is from the power fit of the first six data points of \cite{Zehavi10a} 
in the range $0.25~\hMpc \lesssim r_p \lesssim 2.5~\hMpc$. 
We cut at $r_p \thickapprox 2.5~\hMpc$ because there
is a sharp break at $r_p \thickapprox 2.5 \, \hMpc $ which will be discussed at the end of Section \ref{discuss}.
On the other side, this extension is very robust. In Figure~\ref{fig:zehavi_wp3}, the difference is 
very small among the three extension dashed lines using the first five,
first six and first seven data points of \cite{Zehavi10a}.
Our result (the triangle data points) with correction of fiber collisions fits 
better than the result assuming $p_j = 1$ and $f_j = 1$ (the diamond data points).
We also show the data point before photometry correction Figure~\ref{fig:zehavi_wp3}.


   \subsection{Comparison to previous merger rate results}

Please note that 
the merger time error will be shown in the error bars in the figures from now on.

Our results are consistent with recent measurements
of the merger rates based on counts of close pairs
\citep{Masjedi08a, Patton08a, kitzbichler08a, Bundy09a, De Propris10a, Robaina10a}.
The low dry merger rate an upper limit of $1.8$ $\hpercentGyr$ for massive red 
galaxies (red9) under assumption of $t_{merge} = t_{BT,i}$ here is in good agreement
with a number of other estimates: at $z < 0.36$, \cite{Masjedi08a} obtained an upper 
limit of $1.2$ $\hpercentGyr$ (converted from $1.7$ $\hnewpercentGyr$) for the dry 
merger rate of SDSS LRGs with $M_i < -22.75$;
at $0.45 < z < 0.65$, \cite{De Propris10a} determined a $5 \sigma$ upper limit to the 
dry merger rate of $0.7$ $\hpercentGyr$ (converted from $1.0$ $\hnewpercentGyr$) 
for galaxies with $-23 < M(r)_{k+e,z=0.2} + 5 \log h < -21.5$ 
in the 2dF-SDSS LRG and QSO (2SLAQ) redshift survey.

\cite{Robaina10a} found that the fraction of galaxies ($M > 5 \times 10^{10} \hMsun$)
in pairs separated between 15 and $30 \, \hKpc$ in 3D space is $f_{3Dpair}^{15 - 30 \, \hKpc} = 0.01$
at $z = 0.1$, which is calculate by $(1.0 - 0.3) \times F(z)$, because they find that $30 \sim 40$ percent of galaxies in close pairs have
$r < 15 \, \hKpc$ separations.
They also expect most of the mergers to be majors;
$\ie$, with mass ratio between 1:1 and 1:4.
Our result of $f_{3Dpair}^{15 - 30 \, \hKpc}$ is $0.02$ at $0.03 < z < 0.15$ with mass ratio between 1:1 and 1:4, 
which is consistent with the result of \cite{Robaina10a}.

\begin{figure}
   \centering
   \includegraphics[width=1.\textwidth]{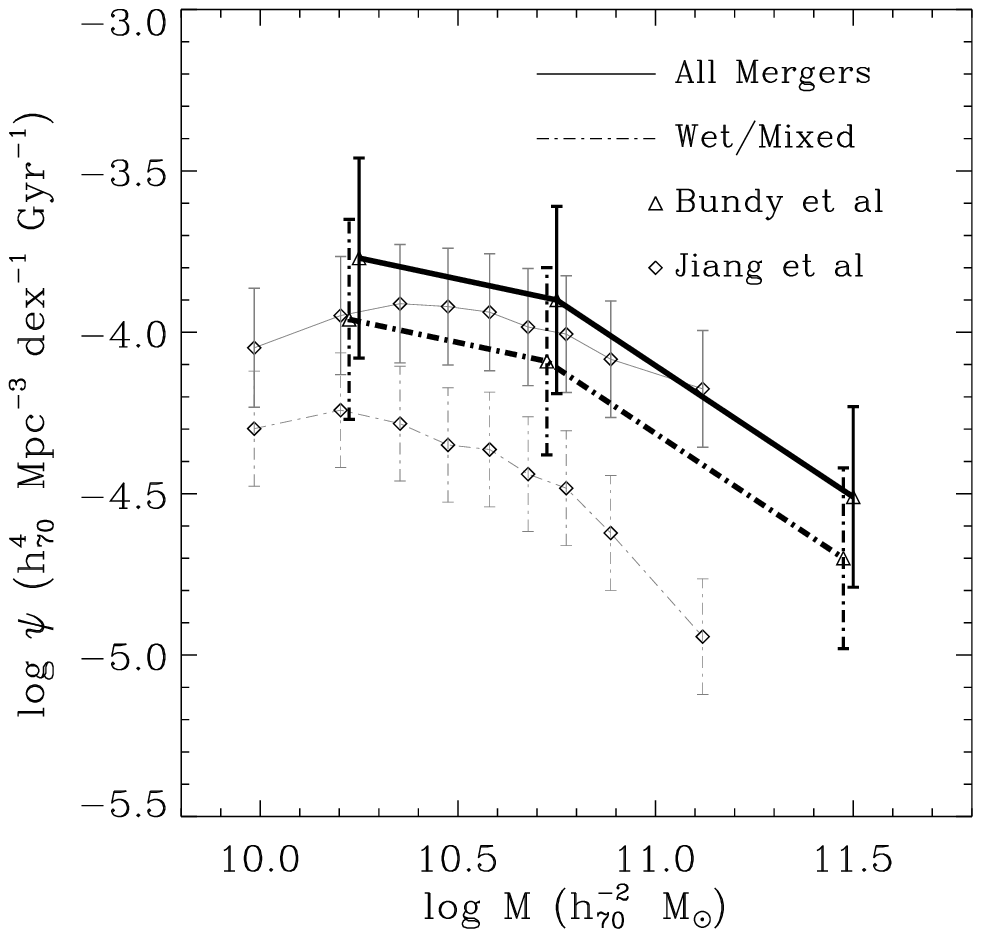}
   \caption{The comparison of our major merger rate MF at $0.03 < z < 0.15$ under assumption of $t_{merge} = t_{KW,i}$ 
with the major merger rate MF of \cite{Bundy09a} at $0.4 < z < 0.7$. Both of our 
mass ratios are $m/M > 0.25$. The open diamonds on the thick lines are our results, and the open triangles on the thin lines are the results of \cite{Bundy09a}. The dashed lines
show the results of excluding the approximate fraction of dry E/S0-E/S0 mergers, and the solid lines indicate the observed merger rate for all galaxies
determined.}
   \label{fig:volume_bundy}
   \end{figure}

With $t_{merge} = t_{KW,i}$ determined, we compute the volumetric merger rate (the number of
mergers per unit time per unit comoving volume) as a function of the
stellar mass of the primary or host galaxy. We call this the {\it
merger rate mass function} (merger rate MF) and denote it using the variable, $\Psi$. 
Figure~\ref{fig:volume_bundy} shows the comparison of our major merger rate MF at $0.03 < z < 0.15$ with the major merger rate MF of \cite{Bundy09a} at $0.4 < z < 0.7$ at mass ratio $m/M > 0.25$.
We can see that both of Figures~\ref{fig:volume_bundy} and ~\ref{fig:bundy} show that our
results are consistent with these previous results.

\begin{figure}
   \centering
   \includegraphics[width=1.\textwidth]{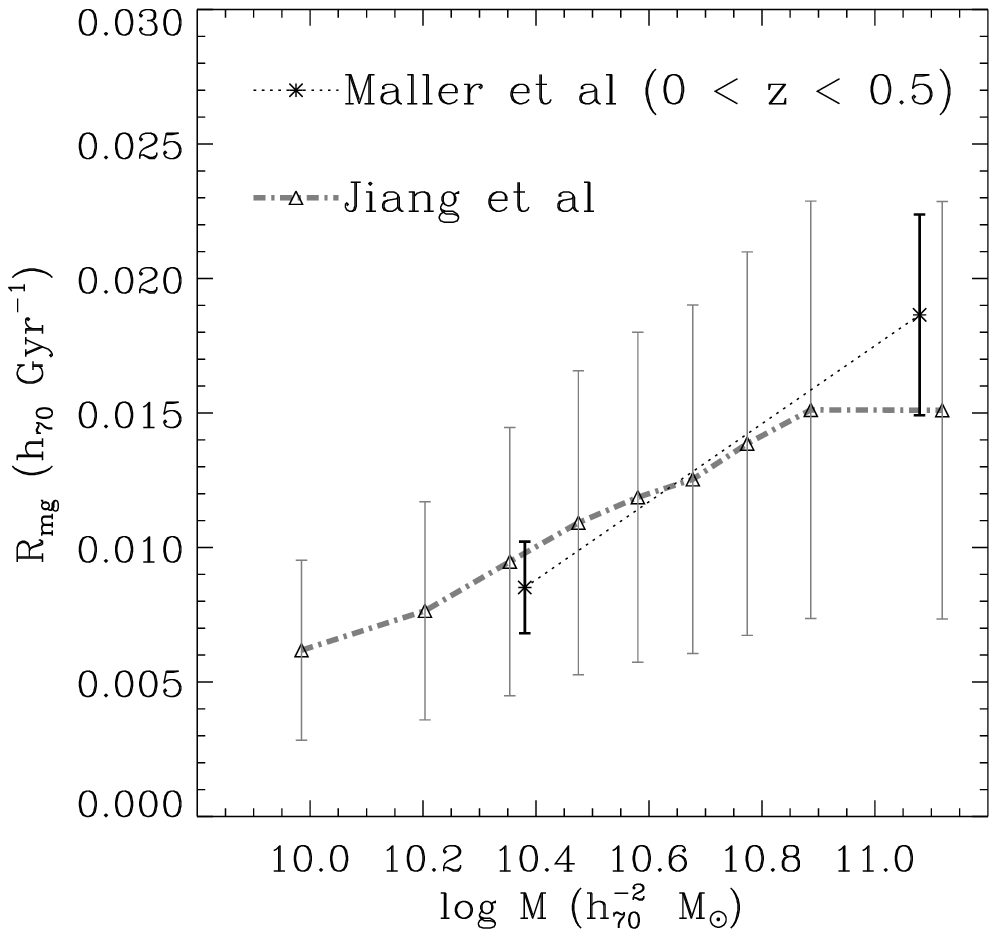}
   \caption{The comparison of our merger rates (per galaxy $\hhGyr$) under assumption of $t_{merge} = t_{KW,i}$ 
at $0.03 < z < 0.15$ with the merger rates of \cite{maller06a} at $0 < z < 0.5$. 
The open triangles on the dashed lines are our results, and the thin dotted lines are the best-fit result 
calculated from equation (5) of \cite{maller06a}. 
Both of our mass ratios are $m/M > 0.5$. Please note that the results of \cite{maller06a} are estimated from 
a flat $\Omega_m= 0.4$ cosmology with $\sigma_8 = 0.8$, a Hubble constant
$H_0 \equiv 100 \, h \, \km \, \s^{-1} \, \Mpc^{-1}$ with $h = 0.65$, a baryon content $\Omega_b = 0.047$, and a spectral index $n=0.93$.
This difference of the two assumptions might lead the difference of the two slopes.}
   \label{fig:maller}
   \end{figure}

\begin{figure}
   \centering
   \includegraphics[width=1.\textwidth]{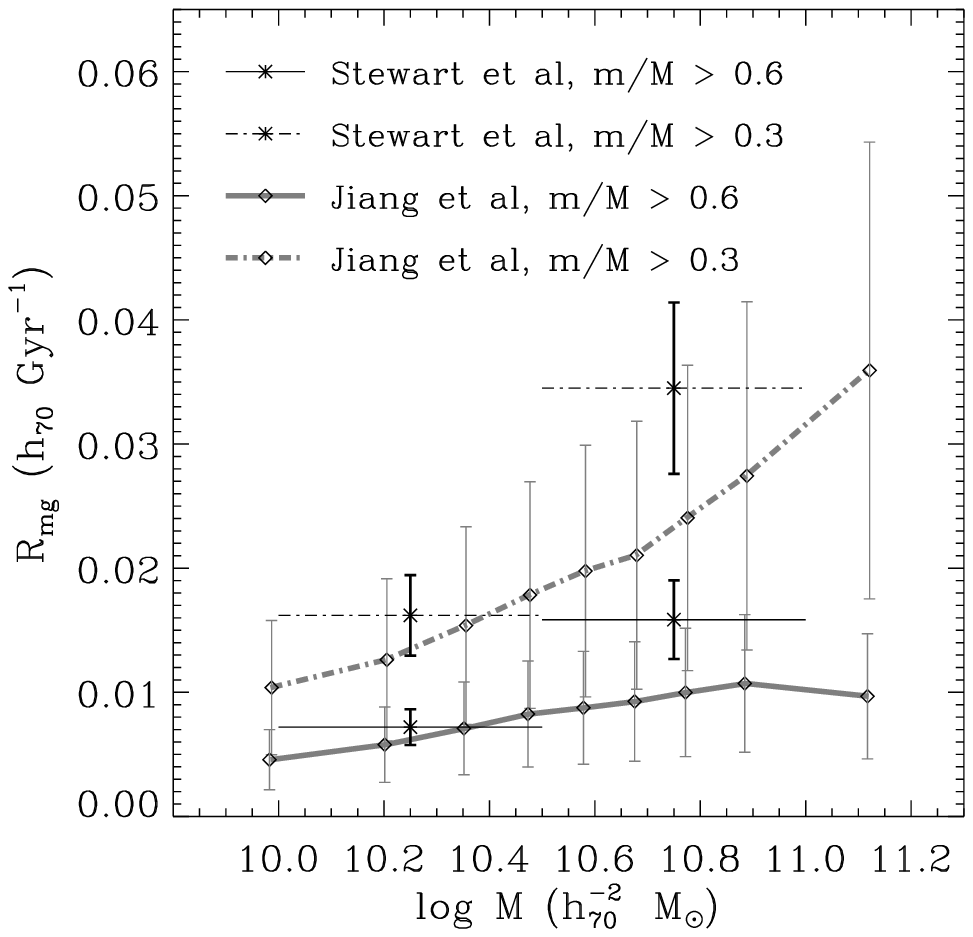}
   \caption{The comparison of our merger rates (per galaxy $\hhGyr$) under assumption of $t_{merge} = t_{KW,i}$ 
at $0.03 < z < 0.15$ with the estimated merger rates calculated by the `Merger Rate 
Fitting Function' of \cite{Stewart09a} at $z = 0.1$. The open diamonds on the thick lines are our results, and the horizontal thin lines are the best-fit results calculated from Table 1 of \cite{Stewart09a}. 
The dashed lines show the results with mass ratios $m/M > 0.3$, and the solid lines show the results with mass ratios $m/M > 0.6$. Please note that the results of \cite{Stewart09a} are estimated from 
a flat $\Omega_m = 1-\Omega_\Lambda = 0.3$ cosmology with a Hubble constant
$H_0 \equiv 100 \, h \, \km \, \s^{-1} \, \Mpc^{-1}$ and $h = 0.7$.}
   \label{fig:stewart}
   \end{figure}

Our results are also consistent with recent merger rates predicted in theories of galaxy
formation in a cosmological context \citep{maller06a, Stewart09a}.
We estimate our merger rate $R_{mg}$ at a certain mass ratio $m/M$ by integrating 
$\Gamma_i$ in equation (\ref{eq:merger}). We compare our $R_{mg}$
with the results of \cite{maller06a} at mass ratio $m/M > 0.5$ (Figure~\ref{fig:maller}) and
the results of \cite{Stewart09a} at mass ratio $m/M > 0.3$ and $m/M > 0.6$ (Figure~\ref{fig:stewart}).
We can see that both of Figures~\ref{fig:maller} and \ref{fig:stewart} show that our
results are consistent with these previous results.

\cite{Wetzel12a} found that from redshift $z=2$ to now, it is around $27$ percent of galaxies similar to our
Milky Way that experienced a merger with mass ratio $m/M > 0.1$, and around $11$ percent
that experienced a merger with mass ratio $m/M > 0.33$.
In our research, the galaxies in blue7 or blue8 subsample are similar to our Milky Way.
If we take our results under assumption of $t_{merge} = t_{KW,i}$ at face value and make the strong assumption
that the growth happens at a non-evolving rate, from redshift $z=2$ to now
(a period of $\approx 10\,\hGyr$), 
we expect the galaxies in blue7 or blue8 subsample to merge by 
$\sim 21$ percent with mass ratio $m/M > 0.1$, and $\sim 10$ percent with mass ratio $m/M > 0.33$,
which is close to the result of \cite{Wetzel12a}.


\section{Discussion}
\label{discuss}

We find that under the assumption of $t_{merge} = t_{KW,i}$, 
the total fractional accretion rates onto red main galaxies are from $[1.3 \pm 0.7]$ to 
$[3.7 \pm 1.9]$ $\hpercentGyr$ depending on stellar mass, and those onto blue main galaxies 
are from $[0.6 \pm 0.3]$ to $[1.1 \pm 0.6]$ $\hpercentGyr$.
We find that at fixed stellar mass, 
the total fractional accretion rates onto red galaxies is greater than that onto blue galaxies. 
The total fractional accretion rate is a stronger function of primary mass for red galaxies 
than that for blue galaxies. We also find that more than $60 \, \unit{percent}$ of the 
total fractional accretion rates are from major mergers with mass ratio between $1:1$ and $1:3$, and less than
$15 \, \unit{percent}$ of the total fractional growth rates are from minor mergers with mass ratio 
between $1:10$ and $1:100$.

The first limitation of the imaging sample arises from the lack of
spectroscopic information on the galaxies from the imaging sample. 
However, for galaxies from imaging sample with stellar masses $M_i > 6 \times 10^9 \, \hMsun$, 
we estimate the real-space number densities $n_i$ from equation (\ref{eq:wpwithoutni}).
So we measured $n_i$, $\rho_{i}$ and $w_{si}(r_p)$ separately. 
But at stellar masses $M_i < 6 \times 10^9 \hMsun$, it is impossible to precisely
measure the real-space number densities $n_i$,
so that we cannot disentangle the clustering power from the number density for 
these small-mass galaxies from the imaging sample, and we only measure the products $n_i \, w_{si}(r_p)$ and $\rho_{i} \, w_{si}(r_p)$
but not either $n_i$, $\rho_{i}$ or $w_{si}(r_p)$ separately.

\begin{figure}
   \centering	
   \includegraphics[width=1.\textwidth]{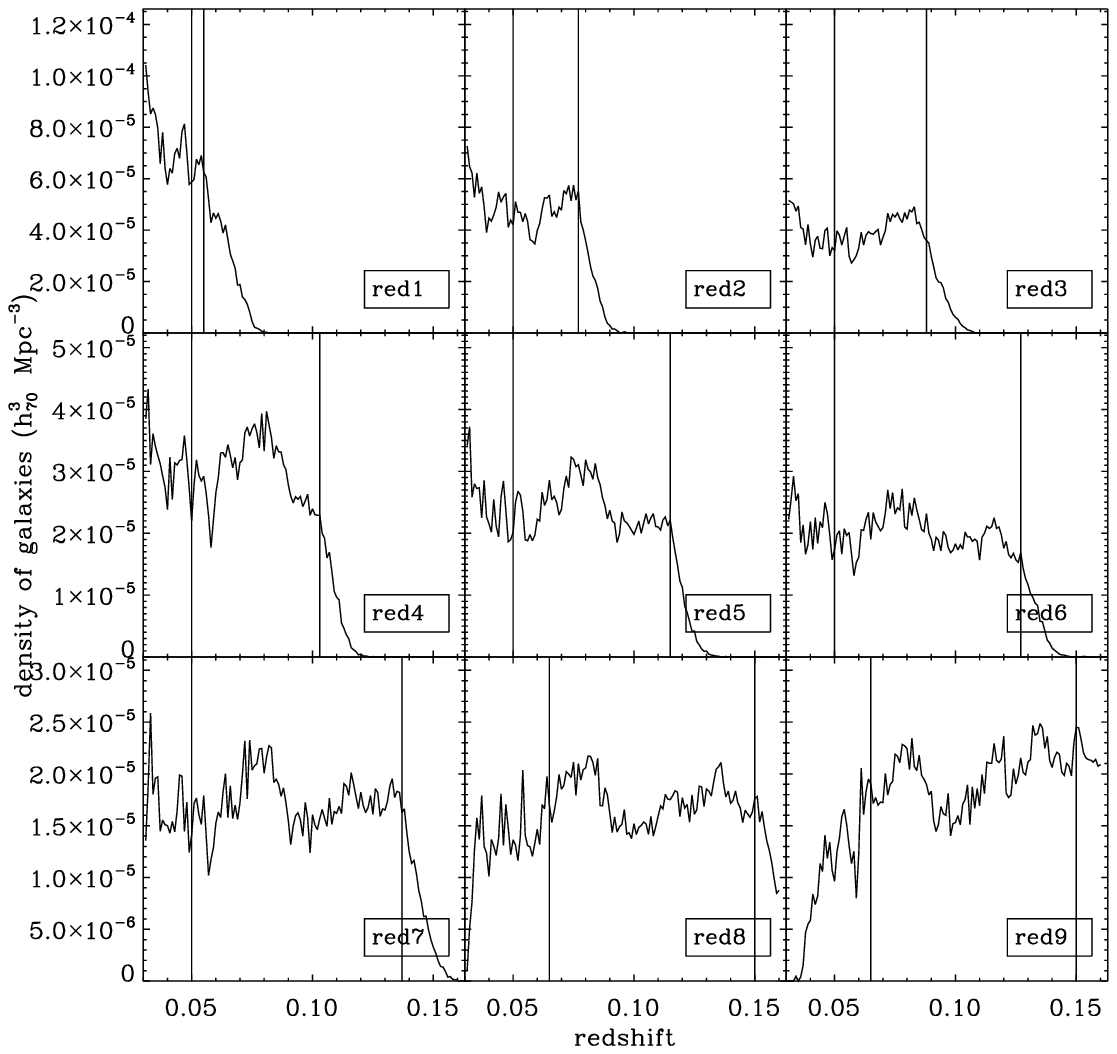}
   \caption{The number densities of each of the 9 red spectroscopic subsamples as a function of redshift. The vertical lines in each graph show the redshift limits used. The number density in red9 appears to rise with redshift, because we have removed galaxies with $r < 14 \, \mag$ (see text).}
   \label{fig:densityred}
   \end{figure}

\begin{figure}
   \centering
   \includegraphics[width=1.\textwidth]{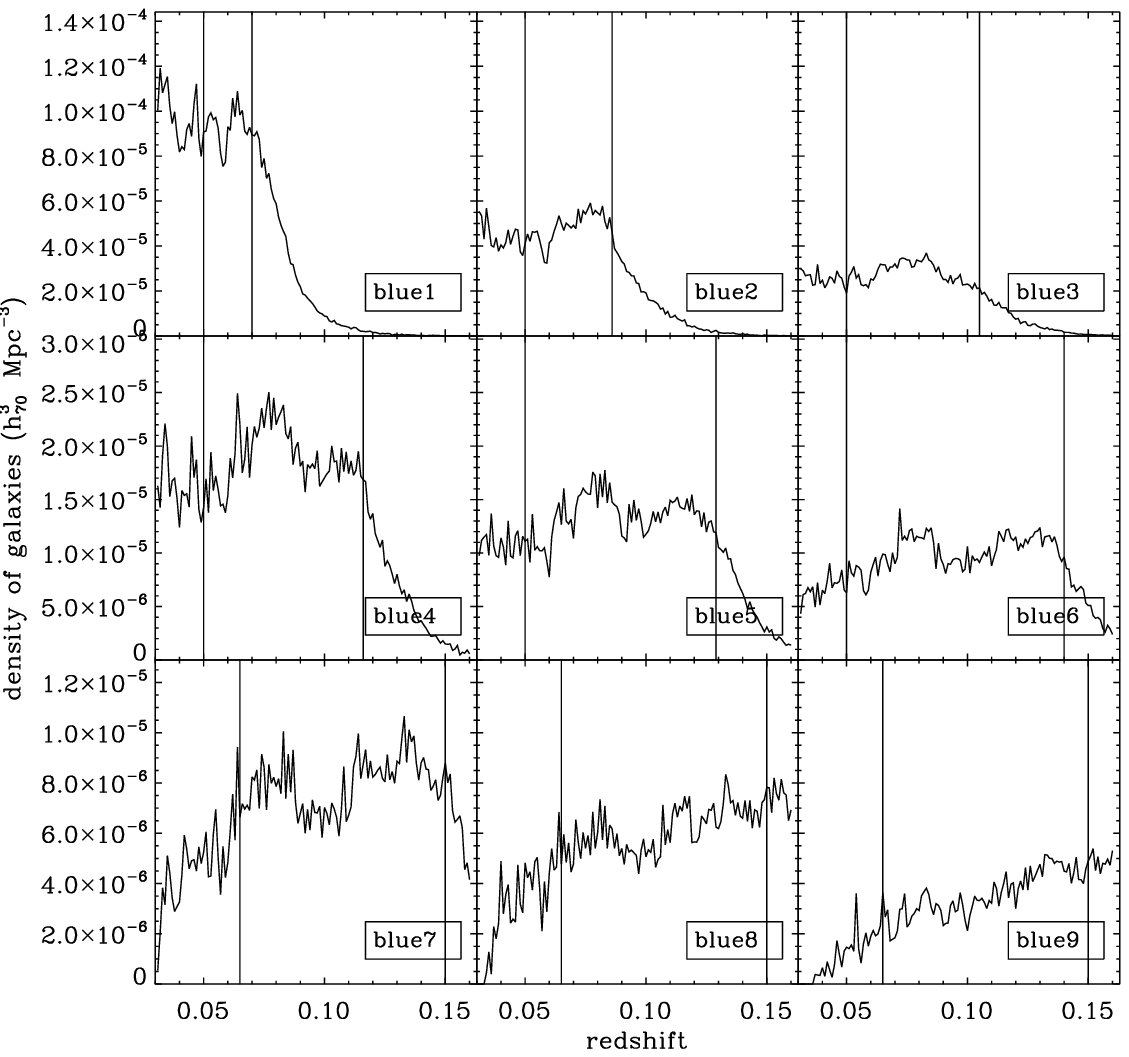}
   \caption{Same as Figure~\ref{fig:densityred}, but for the 9 blue spectroscopic subsamples. The number density in blue8 and blue9 appears to rise with redshift, because we have removed galaxies with $r < 14 \, \mag$ (see text).}.
   \label{fig:densityblue}
   \end{figure}

\begin{figure}
   \centering
   \includegraphics[width=1.\textwidth]{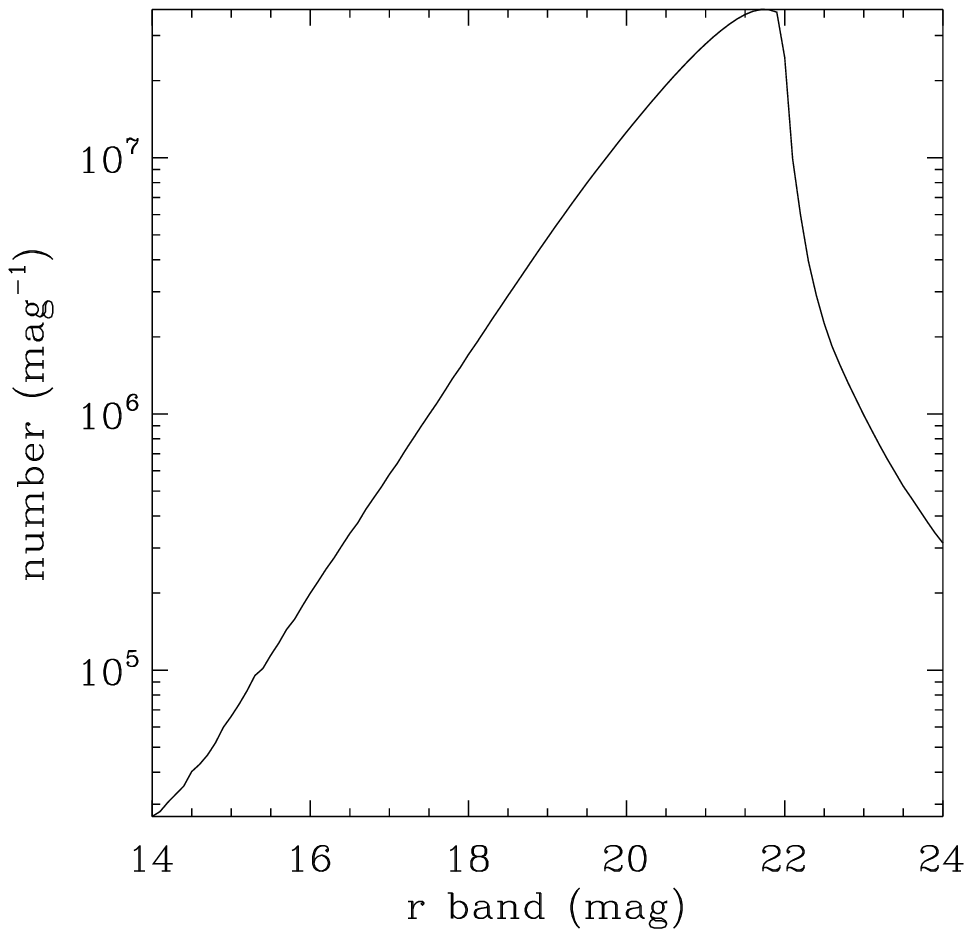}
   \caption{Distribution of apparent $r$ magnitude for the imaging sample. We use only galaxies from the imaging sample with $14 < r < 21.5 \, \mag$.}
   \label{fig:rcut}
   \end{figure}

The second limitation is removing galaxies 
from imaging sample with apparent magnitudes $r > 21.5 \, \mag$
because of the limitation of lack of imaging information on the 
galaxies of SDSS (see Figure~\ref{fig:rcut}).
This cut of apparent magnitudes $r > 21.5 \, \mag$ 
will affect the minor mergers with mass ratio between $1:30$ and $1:100$ 
for the small stellar mass galaxies from spectroscopic samples 
red1$\sim$4 and
blue1$\sim$4 at redshift $z \gtrsim 0.10$, and
the number densities of the galaxies in these spectroscopic samples
decrease sharply 
at redshift $z \gtrsim 0.10$ (see Figures~\ref{fig:densityred} and ~\ref{fig:densityblue}),
which will sharply reduce the effect of this cut.
This cut will cause less than $5 \percent$ error because 
the contribution of the minor mergers with mass ratio between $1:30$ and $1:100$ 
is only $< 5 \percent$.
This assumption is good to take, because it will cause far below $5 \percent$ error.
In order to 
not affect the minor mergers with mass ratio between $1:30$ and $1:100$ at all,
we need our imaging sample to be $\sim 1$ $\mag$ fainter than what we use now.

\begin{figure}
   \centering
   \includegraphics[width=1.\textwidth]{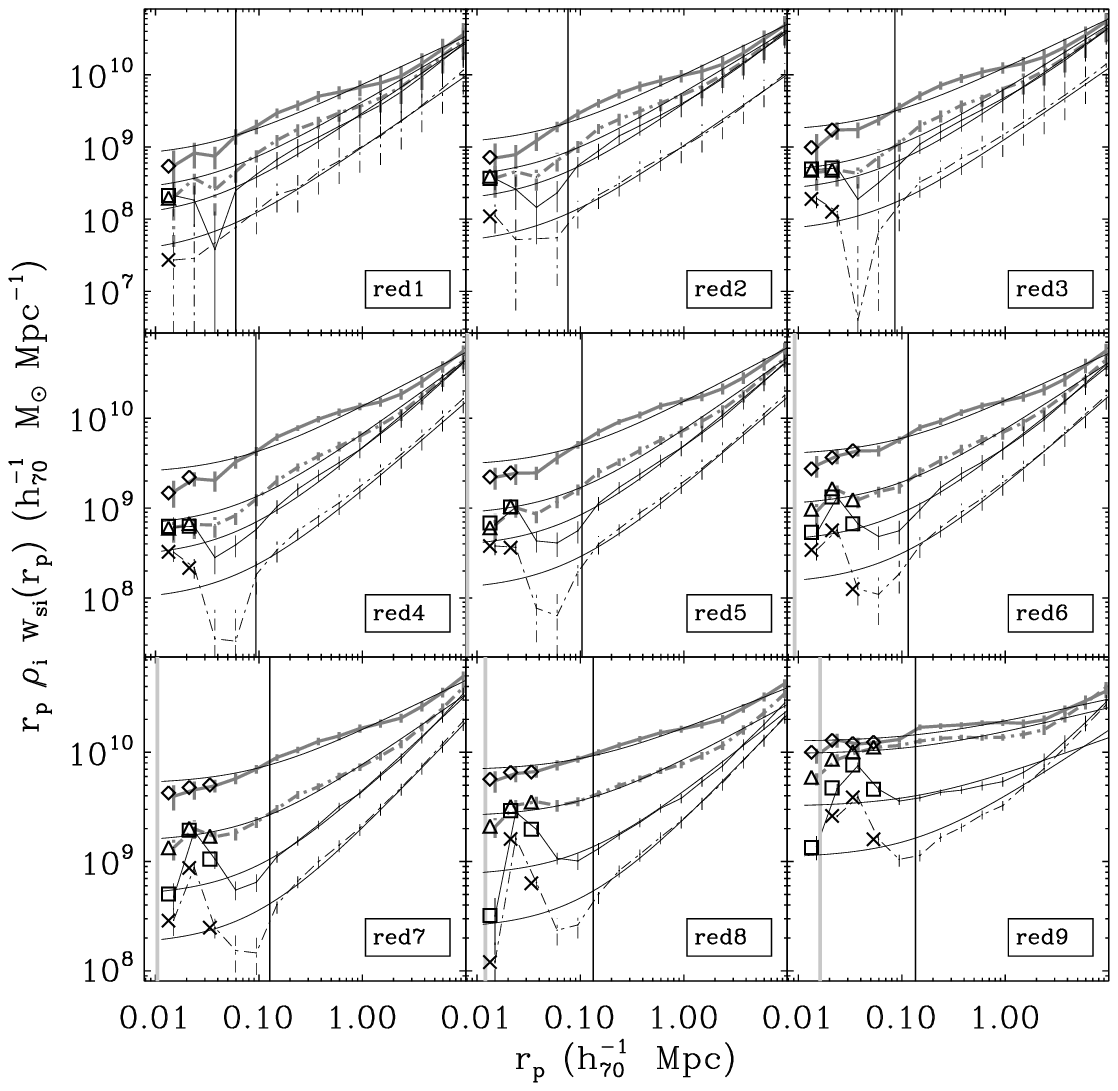}
   \caption{Projected two-dimensional cross-correlation functions $\rho_{i} \, w_{p}(r_p)$ 
between all of the 9 red spectroscopic subsamples and all red 
imaging subsamples, scaled by $r_p$ for better illustration. 
The right vertical thin lines are corresponding to $55$ arcsec at the median redshift
of the spectroscopic galaxies, and the left vertical thick lines are corresponding to 
the median $r_{90}$ for the spectroscopic galaxies.
The error-bars are 
from the jackknife error covariance matrix only. All the lines
show the results after photometry correction: the thick solid line shows 
imaging galaxies with $ 10^{-0.5} < {M_i} / {M_s} < 10^{0} $ (black diamonds are the result before photometry correction), the thick dashed 
line shows $ 10^{-1} < {M_i} / {M_s} < 10^{-0.5} $ (black triangles are the result before photometry correction), the thin solid line shows 
$ 10^{-1.5} < {M_i} / {M_s} < 10^{-1} $ (black squares are the result before photometry correction), and the thin dashed line shows 
$ 10^{-2} < {M_i} / {M_s} < 10^{-1.5} $ (black crosses are the result before photometry correction). The four curves are the fit lines(see text). 
The results before photometry correction
are offset by $12 \percent$ of our interval to the left.}
   \label{fig:nwpredWITHred}
   \end{figure}

\begin{figure}
   \centering
   \includegraphics[width=1.\textwidth]{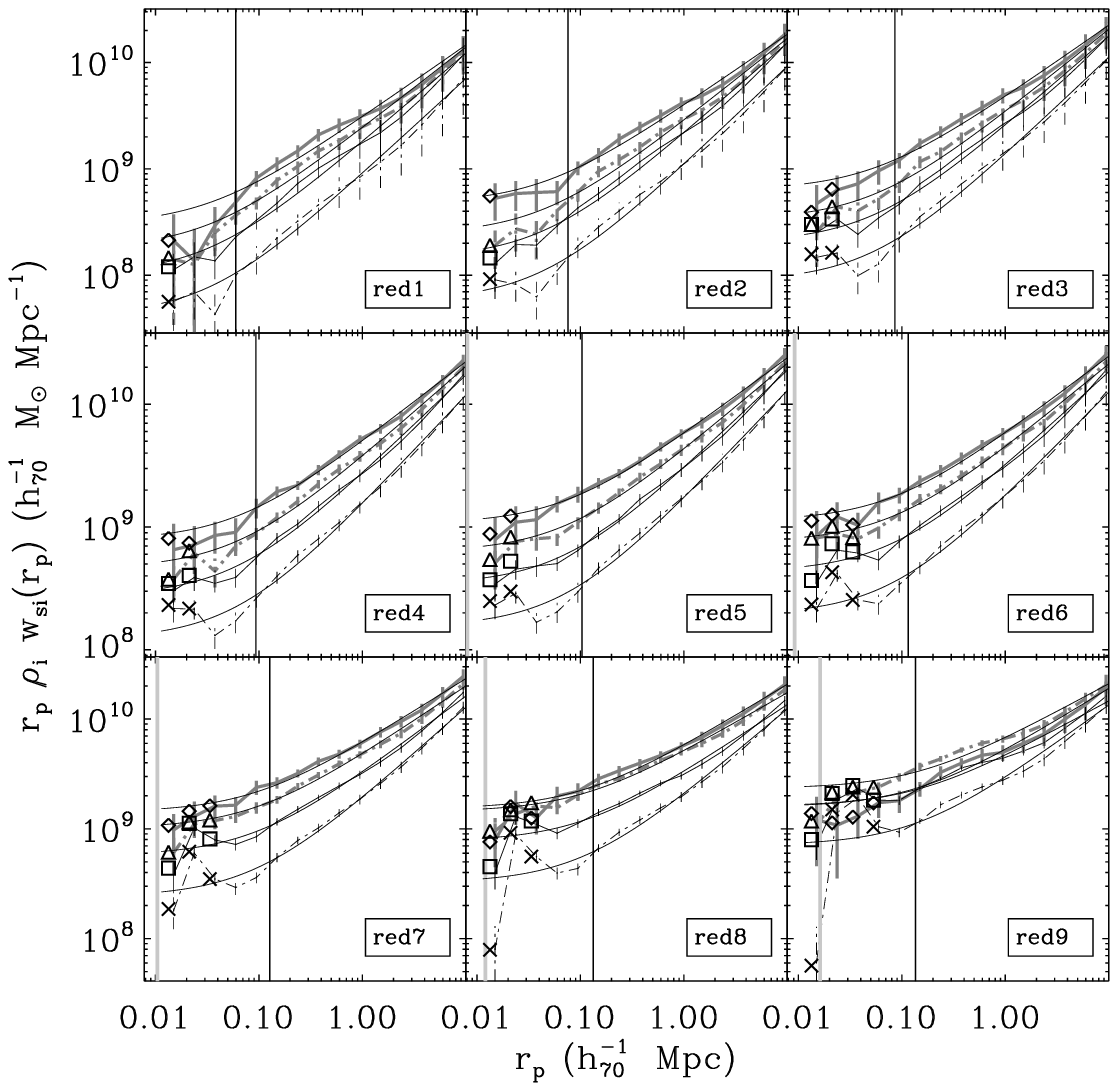}
   \caption{Same as Figure~\ref{fig:nwpredWITHred}, but for the 9 red spectroscopic subsamples cross-correlated with all blue imaging subsamples.}
   \label{fig:nwpredWITHblue}
   \end{figure}

\begin{figure}
   \centering
   \includegraphics[width=1.\textwidth]{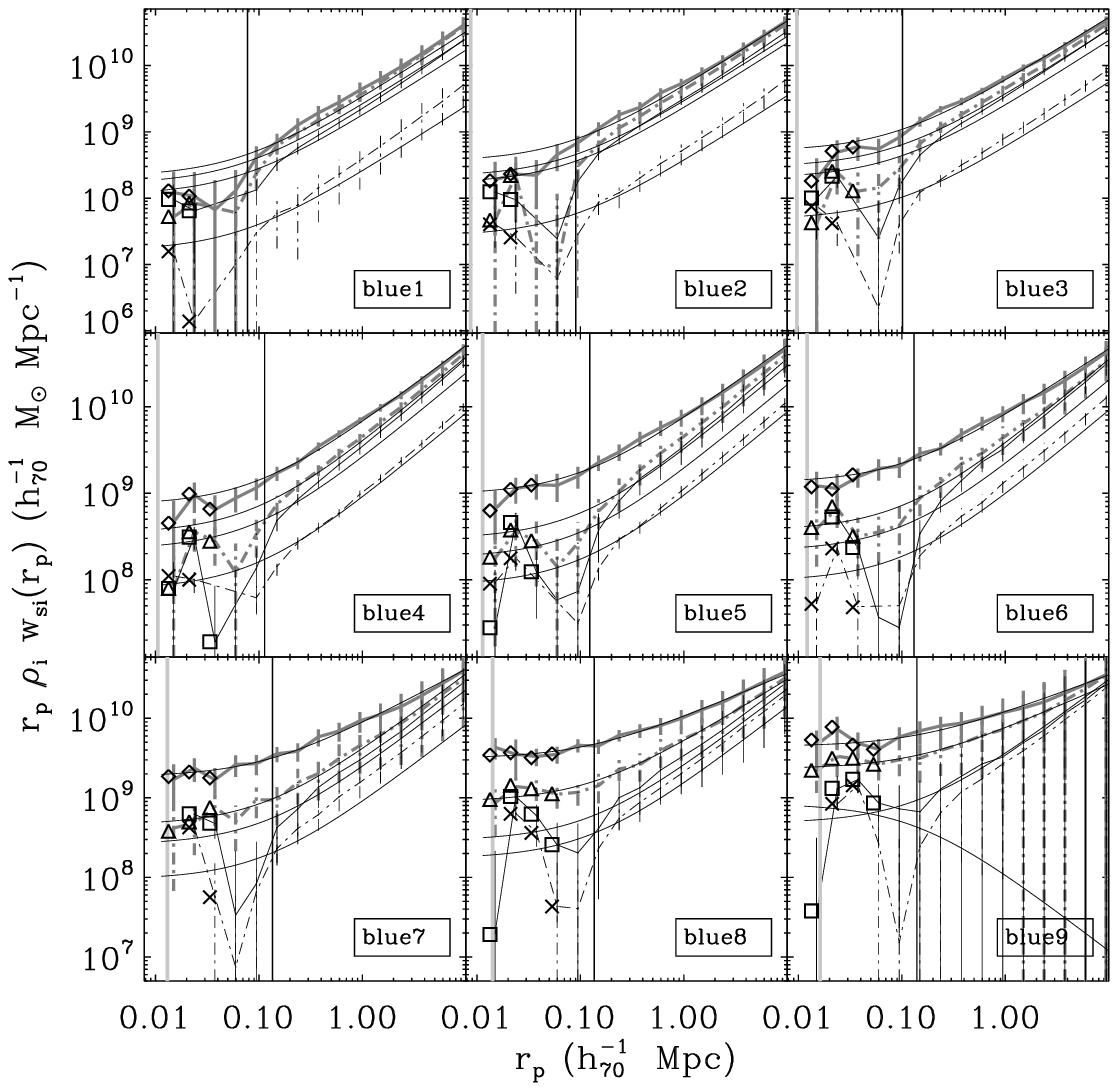}
   \caption{Same as Figure~\ref{fig:nwpredWITHred}, but for the 9 blue spectroscopic subsamples cross-correlated with all red imaging subsamples.}
   \label{fig:nwpblueWITHred}
   \end{figure}

\begin{figure}
   \centering
   \includegraphics[width=1.\textwidth]{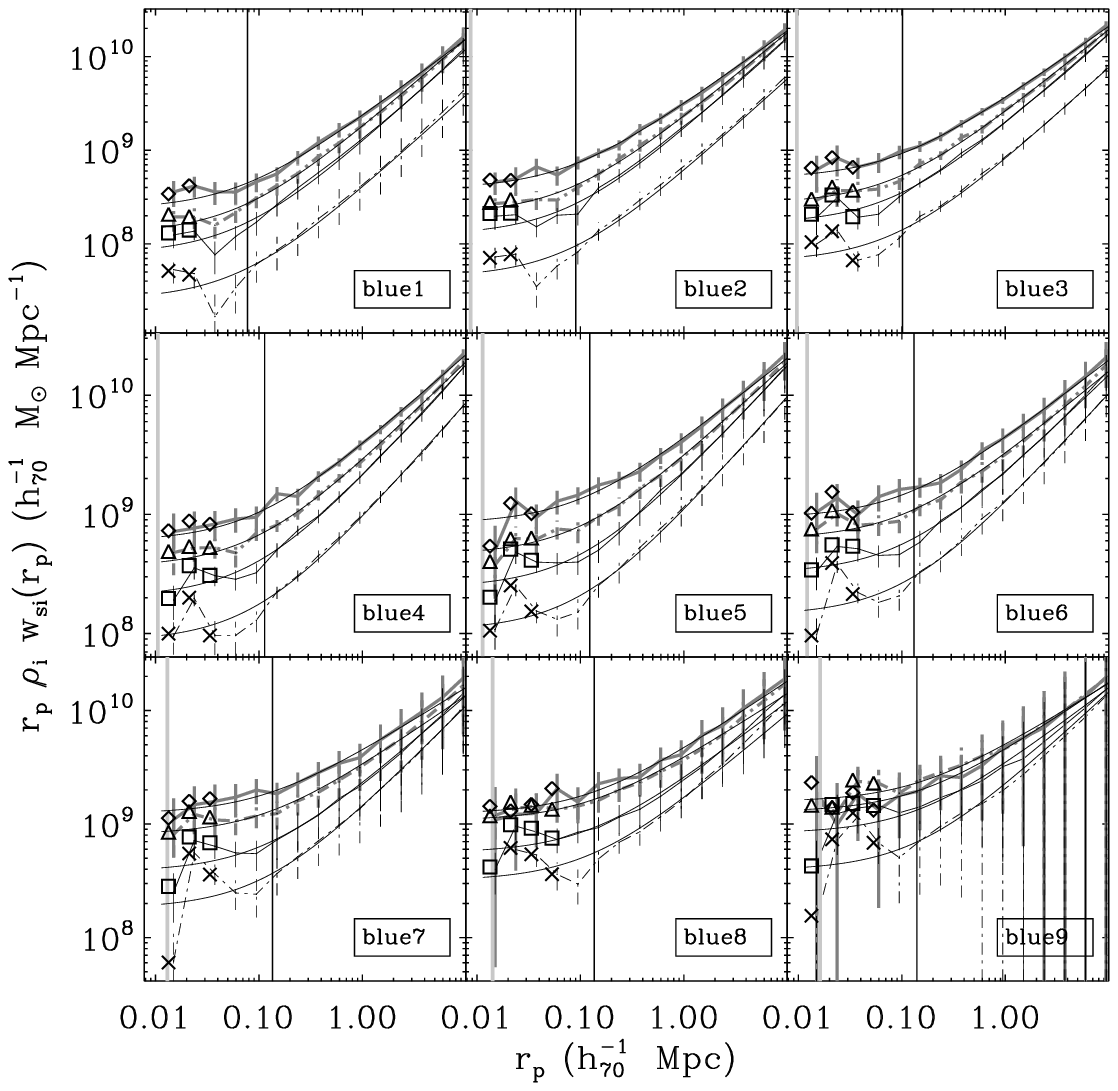}
   \caption{Same as Figure~\ref{fig:nwpredWITHred}, but for the 9 blue spectroscopic subsamples cross-correlated with all blue imaging subsamples.}
   \label{fig:nwpblueWITHblue}
   \end{figure}

We can see `valleys' at the third to fifth data points ($37.6 < r_p < 94.4 \hKpc$) of the two minor-merger curves
($ 10^{-1.5} < {M_i} / {M_s} < 10^{-1} $ and
$ 10^{-2} < {M_i} / {M_s} < 10^{-1.5} $)
in 
each plot of Figures~\ref{fig:nwpredWITHred} through \ref{fig:nwpblueWITHblue}, 
which seems like a kind of issue due to photometric biases or bad deblending. 
However we do not think so, because the photometric biases and bad deblending are very small at 
separations large than $15~\arcs$ (see Table \ref{tab:data0})
and at `valleys' (the third to fifth data points $37.6 < r_p < 94.4 \hKpc$), the photometry correction is nearly zero (below $5 \percent$).
Also, we do not think that the `valleys' will make our conclusions invalid even if that is an issue in minor merger, because 
from the above paragraph we know that only less than
$15 \, \unit{percent}$ of the total fractional accretion rates are from minor mergers with mass ratio between $1:10$ and $1:100$.


If we 
assume
that the growth happens at a non-evolving rate from redshift $z=1$ to now
(a period of $\approx 8\,\hGyr$), we expect the red galaxies to
grow by about $[10 \pm 5] \, \unit{percent}$ to $[28 \pm 14] \, \unit{percent}$ depending on stellar mass under assumption of $t_{merge} = t_{KW,i}$, 
and the red $L^{\ast}$ galaxies (around red7 and red8) grow by about $[20 \pm 10] \, \unit{percent}$.
The 
merger 
rate
may have been different in the past, of course, higher or lower \citep{Lin08a, Chou11a}.
If 
we 
assume that the growth of massive red galaxies ($L^*$ galaxies and above) is mainly from galaxy mergers,
and also assume that
the evolution of the galaxy merger rate per galaxy 
is proportional to $(1+z)^{+3.0 \pm 1.1}$
\citep{lotz11a}
from redshift $z=1$ to now, we expect that the stellar mass density 
of the red massive galaxies ($L^*$ galaxies and above) increased about $\sim 75$ percent under the
assumption of $t_{merge} = t_{KW,i}$, or about $\sim 40$ percent under the
assumption of $t_{merge} = t_{BT,i}$
(see Table \ref{tab:red}),  
which are consistent with recent studies on the high-redshift Universe which find
that the red sequence appears to grow in stellar mass over
time by a factor of $50 \, \unit{percent}$ to $100 \, \unit{percent}$ from redshift $z=1$ to now \citep{bell04a, willmer05a, blanton06a, 
faber07a, conselice07a, Cimatti08a, vanDokkum08a, Saracco09a}.

According to Section \ref{compare},
our results are consistent with the previous clustering results of \cite{Zehavi10a}.
Our estimated merger rates are consistent with the
merger rates estimated by counting of close pairs
\citep{Masjedi08a, Patton08a, kitzbichler08a, Bundy09a, De Propris10a, Robaina10a}.
Our estimated merger rates are also consistent with 
the merger rates predicted in theories of galaxy
formation in a cosmological context \citep{maller06a, Stewart09a, Wetzel12a}.

\begin{figure}
   \centering
   \includegraphics[width=1.\textwidth]{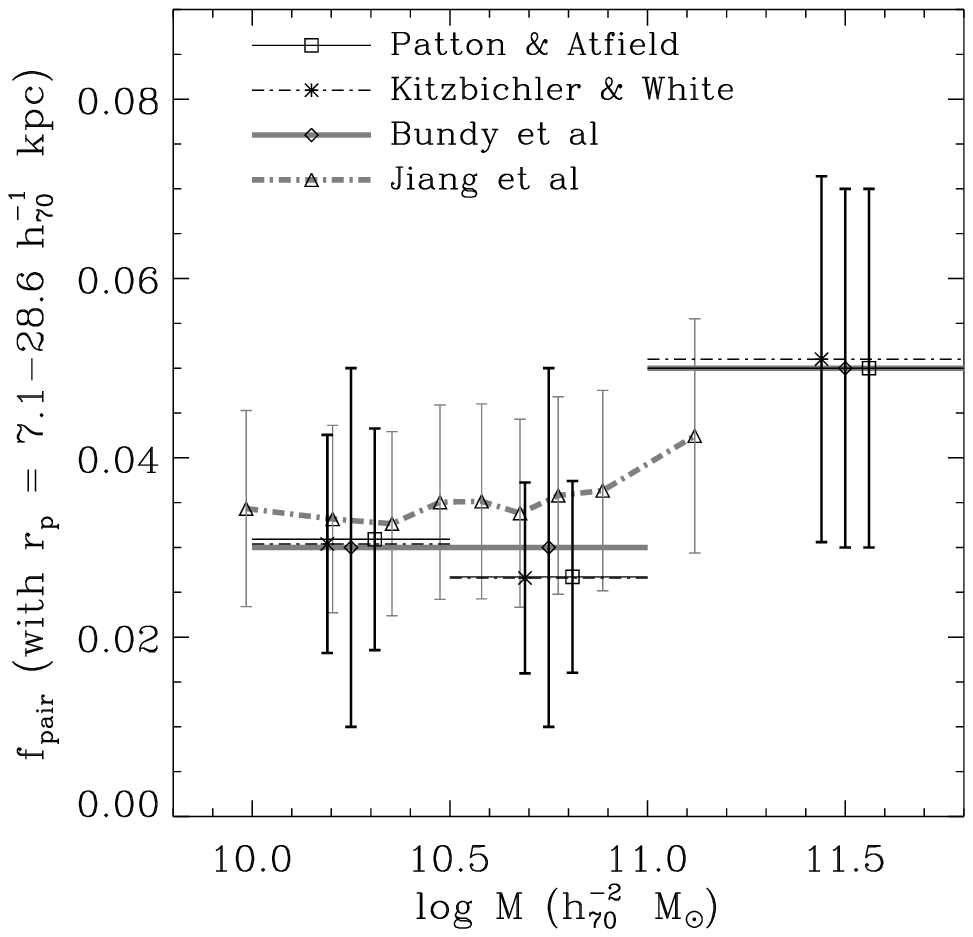}
   \caption{The comparison of our $f_{pair}^{7.1 - 28.6 \, \hKpc}$ at $0.03 < z < 0.15$ with the $f_{pair}^{7.1 - 28.6 \, \hKpc}$ of three recent results
\citep{Patton08a, kitzbichler08a, Bundy09a} at $0.4 < z < 0.7$. 
The open diamonds on the thick solid lines are the results of \cite{Bundy09a}, and the open triangles on the thick dashed lines are our results. The thin horizontal solid lines
show the results of \cite{Patton08a}, and the thin horizontal dashed lines show the results of \cite{kitzbichler08a}. Please note that the $f_{pairs}^{7.1 - 28.6 \, \hKpc}$
of \cite{Patton08a} and \cite{kitzbichler08a} are calculated by $f_{pair}=R_{mg} \times T_{mg} / C_{mg}$.}
   \label{fig:bundy}
   \end{figure}

However, we found that not all merger studies find such low values
when we compared our results with the studies at higher redshift.
The morphological derivations of the merger
fraction (\eg, \citealt{De Propris07a, conselice07a, lotz06a, Lotz08a, Conselice09a}) 
tend to find values of $f_{pair} \approx 0.1$ at $0.4 < z < 1.4$, about $2$
times higher than the results of our pair analysis (see Figure~\ref{fig:bundy}).
The discrepancy can be resolved easily if either
\textsl{(1)}~morphological signatures of merging last for many
dynamical times (\eg, tidal tails)
or \textsl{(2)}~the very minor mergers ($m/M < 0.25$) inflate the merger rates
(please note that our pair analysis in Figure~\ref{fig:bundy} are only estimated from major mergers at
mass ratio $m/M > 0.25$) or \textsl{(3)}~morphological tools for finding mergers maybe 
find some systems which are not involved in mergers or 
\textsl{(4)}~merger rates per galaxy at high redshift may be
larger than those at low redshift.

\begin{figure}
   \centering
   \includegraphics[width=1.\textwidth]{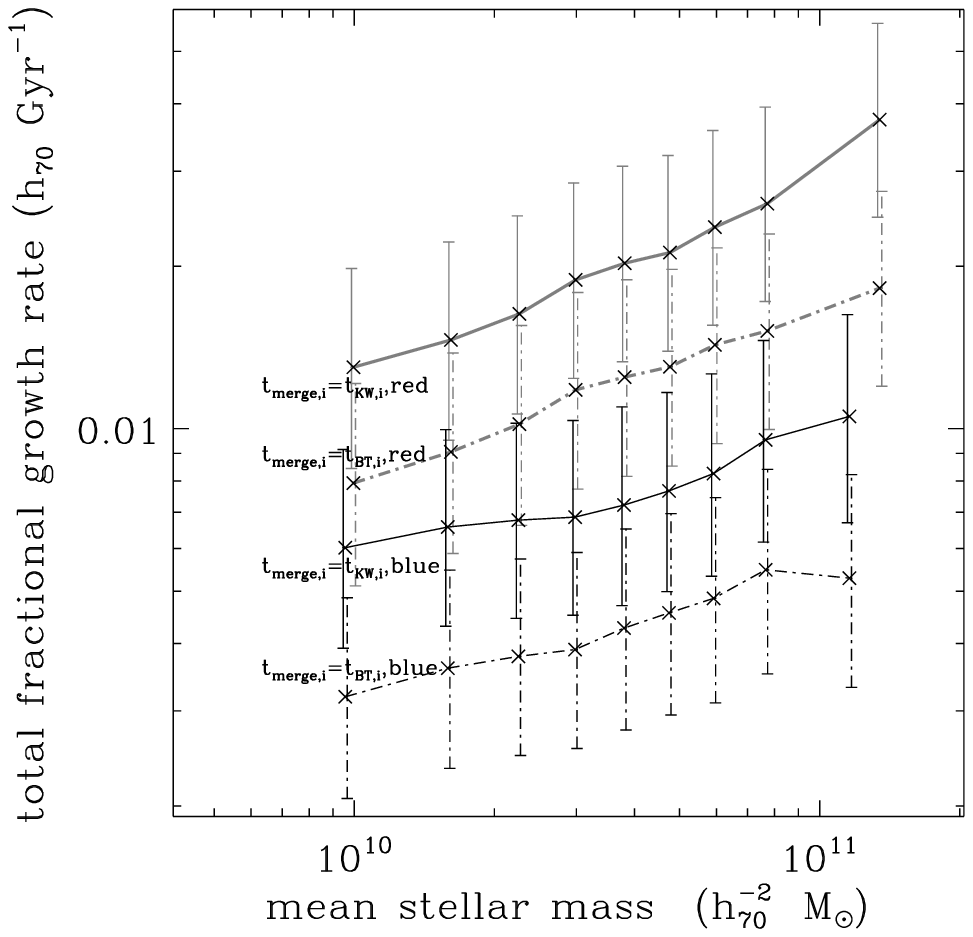}
   \caption{The total fractional accretion rates for each of the 18 spectroscopic subsamples $\hhGyr$ integrating over all the galaxies from the imaging sample. The thick lines are for the 
red spectroscopic subsamples and the thin lines are for the blue spectroscopic subsamples. The solid lines are for the merger rate under assumption of $t_{merge,i} = t_{KW,i}$, and the dashed 
lines are for the merger rate under assumption of $t_{merge,i} = t_{BT,i}$.}
   \label{fig:growth1}
   \end{figure}

The total fractional accretion rates shown in the
solid lines in Figure~\ref{fig:growth1} are upper limits on the true
fractional mass growth. 
There are two reasons: 
\textsl{(1)}~we assume $t_{merge,i} = t_{KW,i}$ for every pair.
However it is not true for close pairs in high velocity dispersion, 
\textsl{(2)}~we assume that the stellar mass growth of the central galaxies from the 
spectroscopic sample is equal to the stellar mass of the galaxy from the imaging sample, see equation (\ref{eq:growth}).
However \cite{lin04a} found that up to $50$ percent of the stars in the galaxies from the imaging sample could be
stripped off before the merger with LRGs is complete.
So
our mass growth rate under the assumption of $t_{merge} = t_{KW,i}$ is an upper limit on the growth by merging.

\begin{figure}
   \centering
   \includegraphics[width=1.\textwidth]{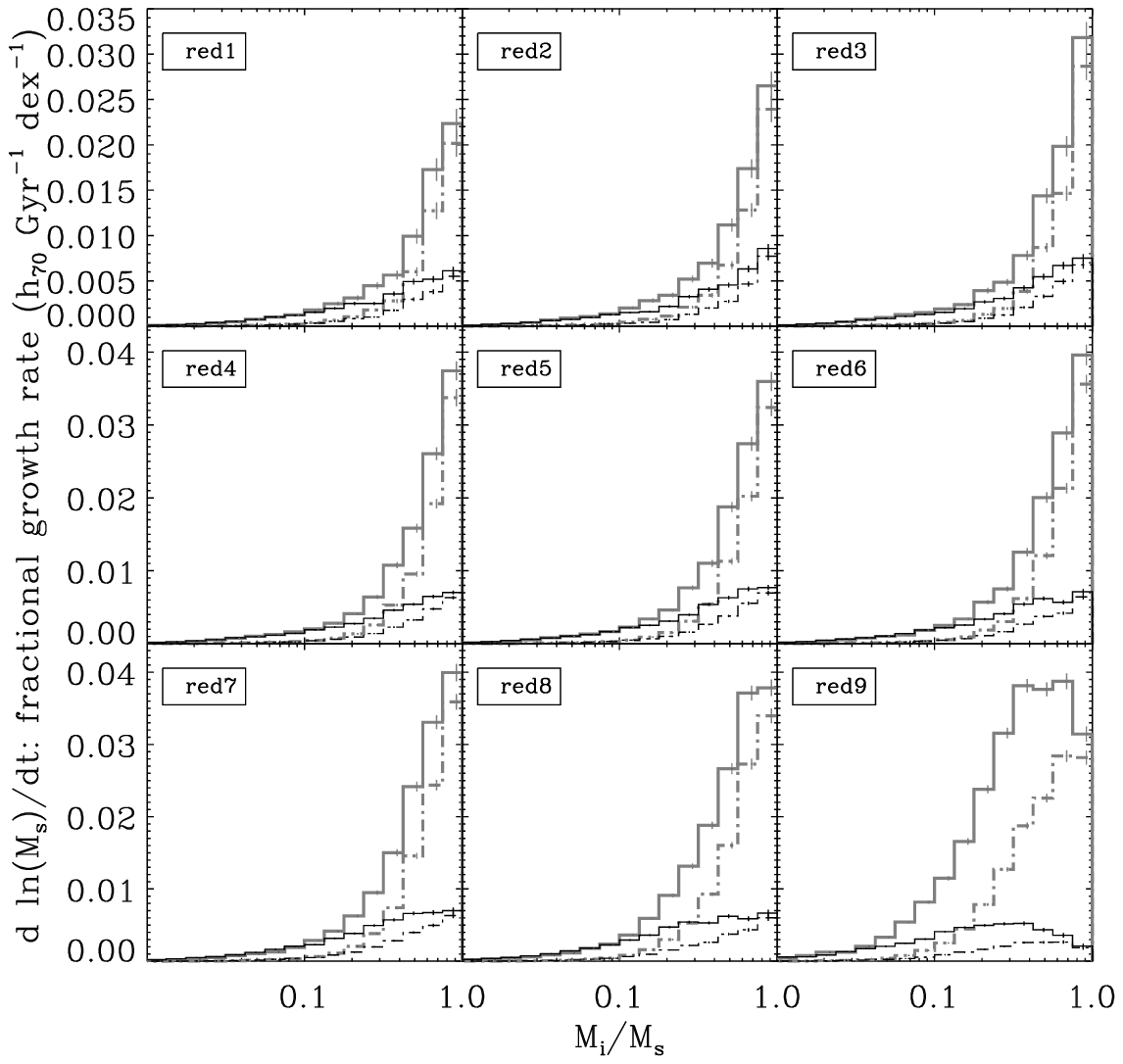}
   \caption{The mean fractional accretion rate for the 9 red spectroscopic subsamples $\hhGyr$ per dex. The thick lines are for mergers with red galaxies from the imaging sample 
and the thin lines are for mergers with blue galaxies from the imaging sample. The solid lines are the merger rate under assumption of $t_{merge,i} = t_{KW,i}$, 
and the dashed lines are for the merger rate under assumption of $t_{merge,i} = t_{BT,i}$.}
   \label{fig:growthred}
   \end{figure}

\begin{figure}
   \centering
   \includegraphics[width=1.\textwidth]{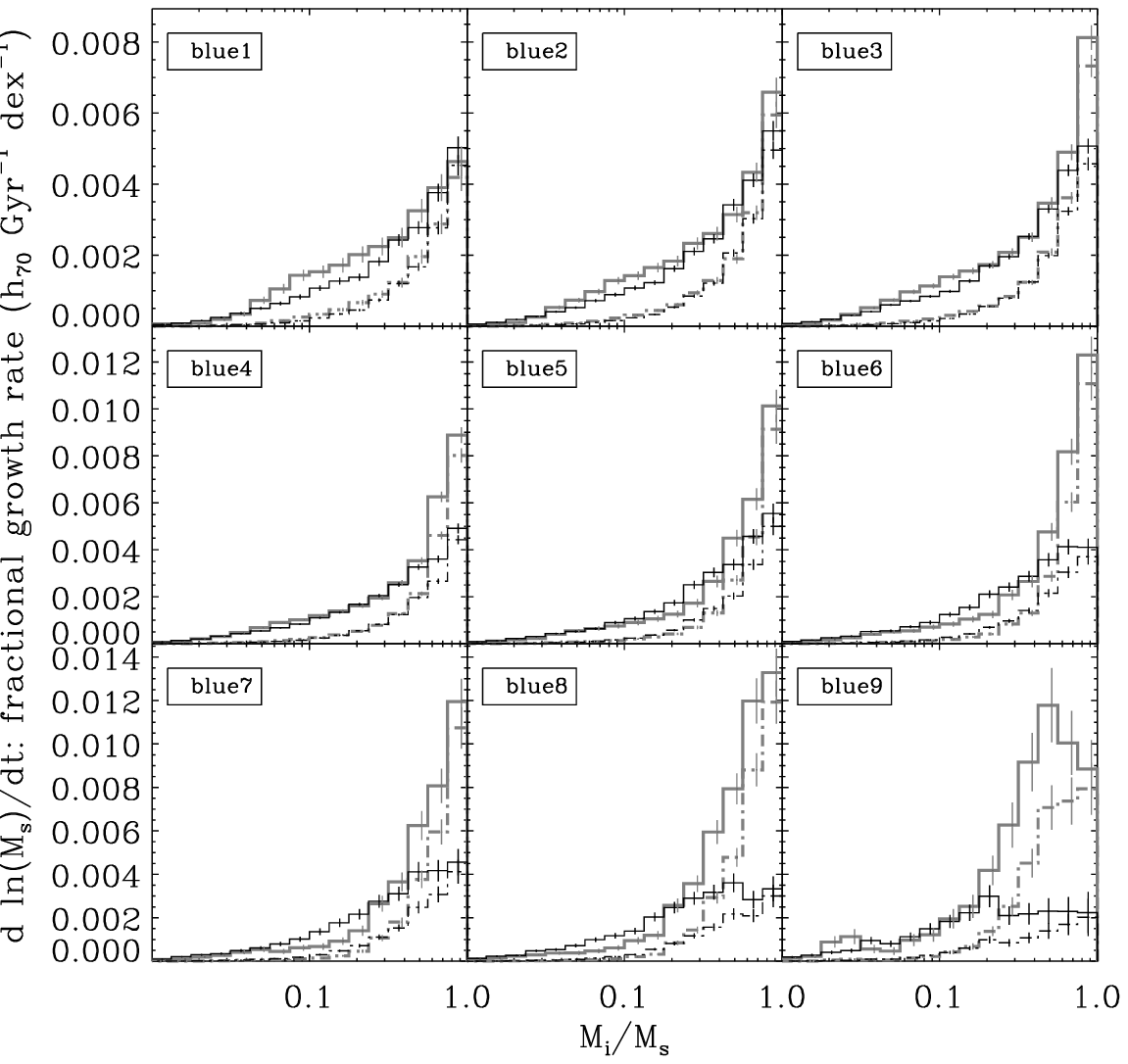}
   \caption{Same as Figure~\ref{fig:growthred}, but for the 9 blue spectroscopic subsamples.}
   \label{fig:growthblue}
   \end{figure}

We find that the accretion onto red and massive blue galaxies
is dominated by mergers with red companions, and that onto small-mass blue galaxies, red and 
blue companions make comparable contributions, this is shown by Table \ref{tab:data1} 
and Figures~\ref{fig:growthred} and \ref{fig:growthblue}. 
So, most of the mass brought into red galaxies by merging is brought by 
``dry mergers'' \citep{bell06b, van Dokkum05a, Masjedi08a}.

We find that all the contributions to growth decrease with decreasing 
stellar mass at the small-mass end for all of 18
spectroscopic subsamples.
The contribution to growth decreases with
decreasing $M_i/M_s$ since $M_i/M_s < 0.4$.
For all 18 subsamples, the curves essentially decrease to zero
by $M_i/M_s < 0.01$ for $t_{merge} = t_{KW,i}$ and by $M_i/M_s < 0.1$ for $t_{merge} = t_{BT,i}$, so calculation of the total amount of mass
brought in by merger activities does not require consideration of
galaxies from the imaging sample with $M_i/M_s < 0.01$.

\begin{figure}
   \centering
   \includegraphics[width=1.\textwidth]{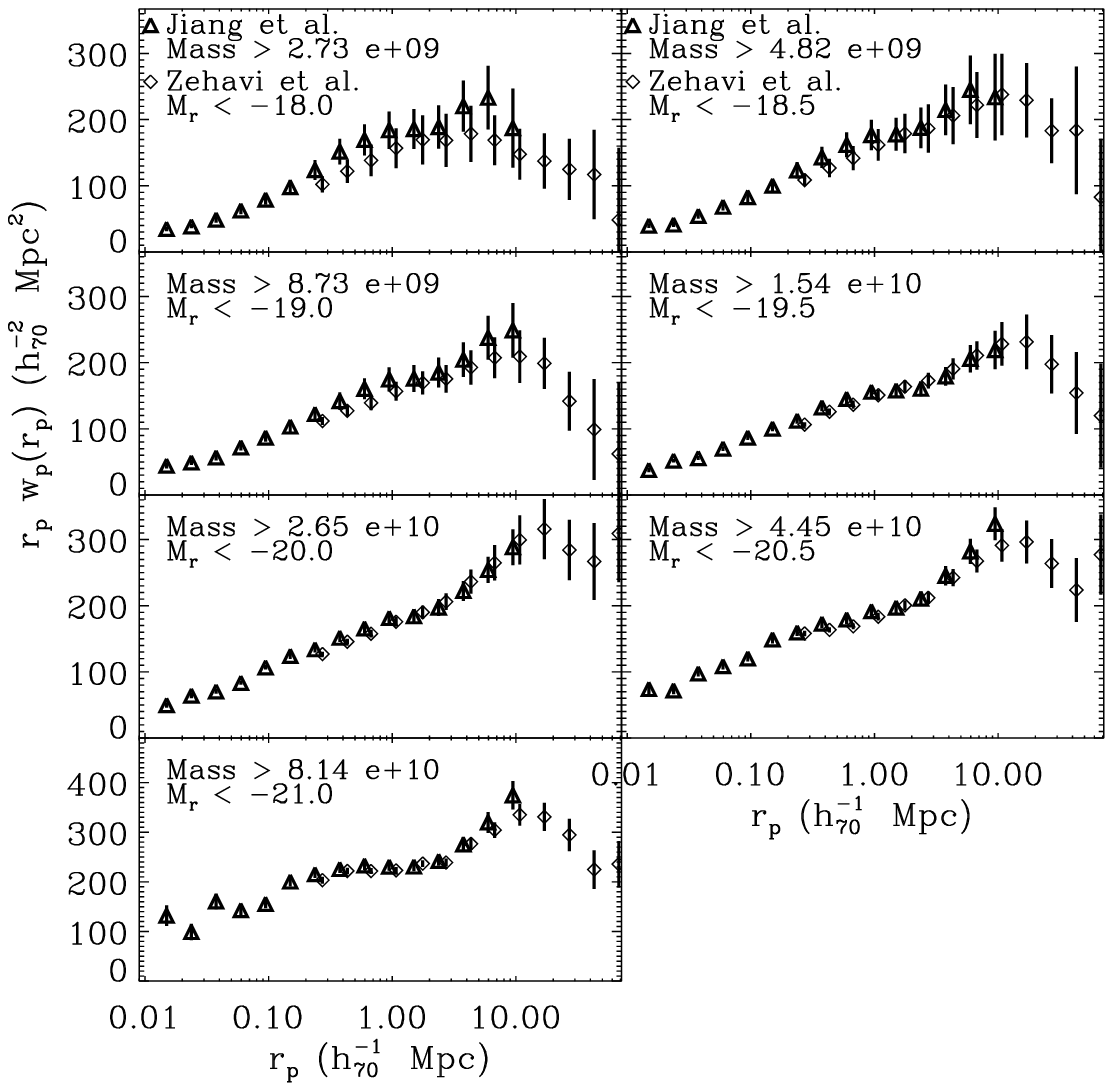}
   \caption{Projected correlation function $w_p(r_p)$ for the spectroscopic subsamples corresponding to mass-threshold samples as labeled, calculated as described in the text on small scales, combined with projected correlation function on intermediate scales from \cite{Zehavi10a}. Please note that in order to
compare these results easily, we offset the points of \cite{Zehavi10a} by $12 \percent$ of our interval  to the right.}
   \label{fig:zehavi_wp}
   \end{figure}

\begin{figure}
   \centering
   \includegraphics[width=1.\textwidth]{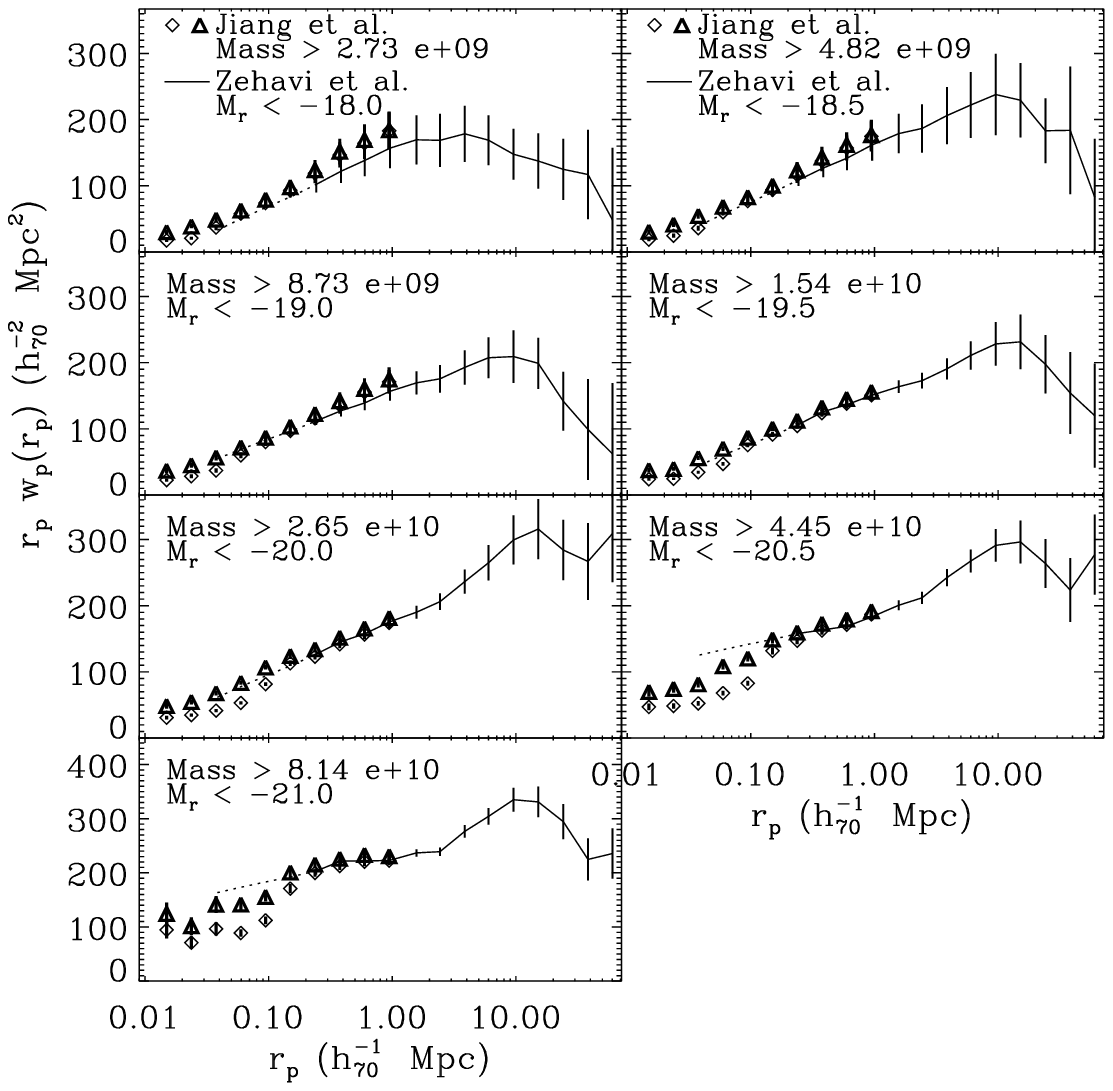}
   \caption{Projected correlation function $w_p(r_p)$ similar as Figure~\ref{fig:zehavi_wp}. There is no offset in this graph.
The triangle points are our $w_p(r_p)$ with correction of fiber collisions, the diamond points are 
the $w_p(r_p)$ assuming $p_j = 1$ and $f_j = 1$, the solid lines with thin error bars are
the $w_p(r_p)$ of \cite{Zehavi10a} and the dashed lines are the extension lines of \cite{Zehavi10a} described in the text. 
Please note that in order to compare these results easily, we only display our first ten data points.}
   \label{fig:zehavi_wp2}
   \end{figure}

\begin{figure}
   \centering
   \includegraphics[width=1.\textwidth]{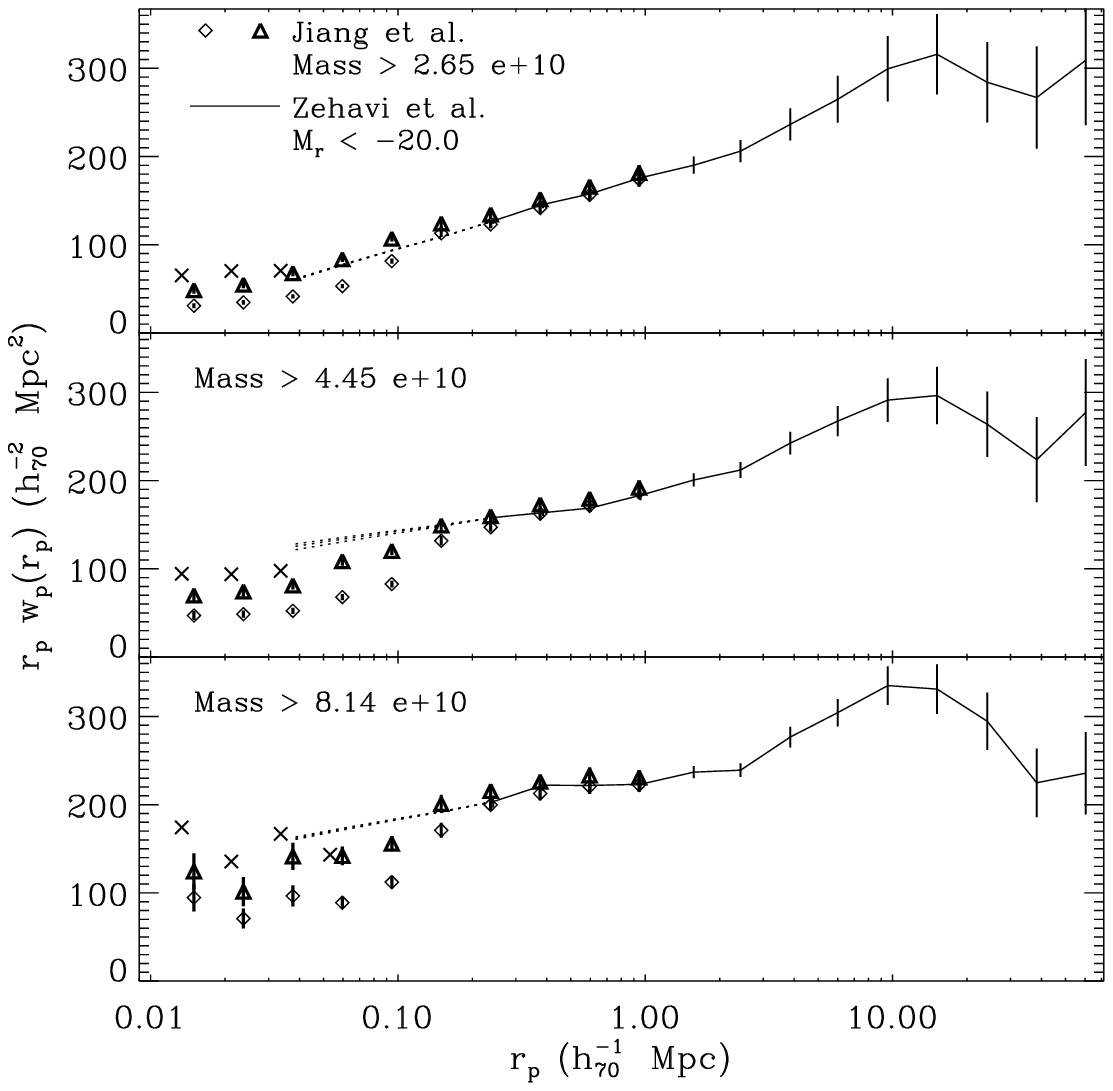}
   \caption{Projected correlation function $w_p(r_p)$ similar as Figure~\ref{fig:zehavi_wp2}, but only display the last three plots.
The black crosses in the second and third plots are our result before photometry correction which are offseted by $12 \percent$ of our interval  to the left.
Please note that the three extension dashed lines are fitting from the first five,
first six and first seven data points of \cite{Zehavi10a}.
}
   \label{fig:zehavi_wp3}
   \end{figure}

From Figures~\ref{fig:nwpredWITHred} through \ref{fig:nwpblueWITHblue} and Figure~\ref{fig:zehavi_wp},
we find a sharp break at $r_p \thickapprox 2.5 \, \hMpc $ and a less-sharp transition 
at $r_p \thickapprox 0.43 \, \hMpc $.
These two transitions are also found in LRGs by \cite{Masjedi08a},
which
can be explained in the context of the ``halo occupation'' picture of galaxy clustering
\citep{peacock00a, scoccimarro01a, berlind02a, cooray02a, zheng05a, watson10a, watson12a}:
\textsl{(1)}~the mergers at $r_p < 0.43 \, \hMpc$ are the one-halo mergers (both of the two merging galaxies
are inside one halo);
\textsl{(2)}~the mergers at $r_p > 2.5 \, \hMpc$ are the two-halo mergers 
(the two merging galaxies are separately inside two nearby halos);
\textsl{(3)}~the mergers at $0.43 < r_p < 2.5 \, \hMpc$ 
are the mixed-halo mergers (some of the mergers are the one-halo mergers, the others are the two-halo mergers).
So at $r_p \thickapprox 0.43 \, \hMpc$ the mergers transfer from the one-halo mergers to
the mixed-halo mergers, and at $r_p \thickapprox 2.5 \, \hMpc$ the mergers transfer from 
the mixed-halo mergers to the two-halo mergers.
These two transitions are clearer in 
Figure~\ref{fig:nwpredWITHred} of mergers between two red galaxies than that in the other 
three figures.

It is a pleasure to thank Douglas Watson, Jeremy Tinker
and our anonymous
referee for valuable input.

 This research made use of public SDSS data.
 Funding for the SDSS and SDSS-II has been provided by
 the Alfred P. Sloan Foundation,
 the Participating Institutions,
 the National Science Foundation,
 the U.S. Department of Energy,
 the National Aeronautics and Space Administration,
 the Japanese Monbukagakusho,
 the Max Planck Society,
 and the Higher Education Funding Council for England.
 The SDSS Web Site is http://www.sdss.org/.

 SDSS is managed by
 the Astrophysical Research Consortium for the Participating Institutions.
 The Participating Institutions are
 the American Museum of Natural History,
 Astrophysical Institute Potsdam,
 University of Basel,
 University of Cambridge,
 Case Western Reserve University,
 University of Chicago,
 Drexel University,
 Fermilab,
 the Institute for Advanced Study,
 the Japan Participation Group,
 Johns Hopkins University,
 the Joint Institute for Nuclear Astrophysics,
 the Kavli Institute for Particle Astrophysics and Cosmology,
 the Korean Scientist Group,
 the Chinese Academy of Sciences (LAMOST),
 Los Alamos National Laboratory,
 the Max-Planck-Institute for Astronomy (MPIA),
 the Max-Planck-Institute for Astrophysics (MPA),
 New Mexico State University,
 Ohio State University, University of Pittsburgh,
 University of Portsmouth,
 Princeton University,
 the United States Naval Observatory,
 and the University of Washington.



\newpage
\clearpage
\begin{center}
\begin{deluxetable}{cccccccc}
\tablecaption{Spectroscopic Subsamples\label{tab:data0}}
\tablecolumns{7}
\tablehead{
\colhead{Subsample} & $\left<M\right>$\tablenotemark{\protect{\ref{1mass}}} & \colhead{number} & $\left<z\right>$\tablenotemark{\protect{\ref{1zmean}}} & $z_{min}$\tablenotemark{\protect{\ref{1zmin}}} & $z_{max}$\tablenotemark{\protect{\ref{1zmax}}} & $\sigma_{v}$ or $V_{c} $\tablenotemark{\protect{\ref{1velocity}}} & angular separation\tablenotemark{\protect{\ref{1recover}}}\\
\colhead{}  & [$10^{10}\hMsun$] & \colhead{} & \colhead{} & \colhead{} & \colhead{} & [$\km \, \s^{-1}$] & [arcsec] 
}

\startdata
red0   & $    $ & $       $ & $       $ & $      $ & $       $ & $    $  & $               $\\
red1   & $1.00$ & $ 11115 $ & $ 0.052 $ & $ 0.05 $ & $ 0.055 $ & $88.8$  & $35.0$ and $55.4$\\
red2   & $1.61$ & $ 17023 $ & $ 0.064 $ & $ 0.05 $ & $ 0.077 $ & $105 $  & $28.5$ and $45.1$\\
red3   & $2.26$ & $ 20585 $ & $ 0.073 $ & $ 0.05 $ & $ 0.088 $ & $119 $  & $25.2$ and $39.9$\\
red4   & $2.99$ & $ 22904 $ & $ 0.081 $ & $ 0.05 $ & $ 0.103 $ & $131 $  & $22.7$ and $36.0$\\
red5   & $3.81$ & $ 25184 $ & $ 0.090 $ & $ 0.05 $ & $ 0.115 $ & $143 $  & $20.6$ and $32.6$\\
red6   & $4.77$ & $ 27809 $ & $ 0.099 $ & $ 0.05 $ & $ 0.127 $ & $155 $  & $18.8$ and $29.8$\\
red7   & $5.95$ & $ 31390 $ & $ 0.108 $ & $ 0.05 $ & $ 0.137 $ & $167 $  & $17.1$ and $27.2$\\
red8   & $7.72$ & $ 34919 $ & $ 0.113 $ & $ 0.05 $ & $ 0.150 $ & $183 $  & $16.4$ and $26.0$\\
red9   & $13.5$ & $ 40036 $ & $ 0.116 $ & $ 0.05 $ & $ 0.150 $ & $223 $  & $16.0$ and $25.4$\\
blue0  & $    $ & $       $ & $       $ & $      $ & $       $ & $    $  & $    $     $    $\\
blue1  & $0.96$ & $ 36470 $ & $ 0.069 $ & $ 0.05 $ & $ 0.070 $ & $123 $  & $26.6$ and $42.2$\\
blue2  & $1.59$ & $ 30562 $ & $ 0.080 $ & $ 0.05 $ & $ 0.086 $ & $137 $  & $22.9$ and $36.3$\\
blue3  & $2.25$ & $ 26992 $ & $ 0.090 $ & $ 0.05 $ & $ 0.105 $ & $148 $  & $20.5$ and $32.6$\\
blue4  & $2.98$ & $ 24688 $ & $ 0.099 $ & $ 0.05 $ & $ 0.116 $ & $157 $  & $18.8$ and $29.8$\\
blue5  & $3.80$ & $ 22404 $ & $ 0.106 $ & $ 0.05 $ & $ 0.129 $ & $166 $  & $17.5$ and $27.7$\\
blue6  & $4.75$ & $ 19774 $ & $ 0.112 $ & $ 0.05 $ & $ 0.140 $ & $175 $  & $16.6$ and $26.3$\\
blue7  & $5.92$ & $ 16197 $ & $ 0.115 $ & $ 0.05 $ & $ 0.150 $ & $184 $  & $16.1$ and $25.5$\\
blue8  & $7.65$ & $ 12667 $ & $ 0.117 $ & $ 0.05 $ & $ 0.150 $ & $197 $  & $16.0$ and $25.3$\\
blue9  & $11.6$ & $  7551 $ & $ 0.120 $ & $ 0.05 $ & $ 0.150 $ & $219 $  & $15.6$ and $24.6$\\

\enddata

\tablecomments{Information for the galaxies in the spectroscopic subsamples.}

\setcounter{tableone}{1}
\makeatletter
\let\@currentlabel\oldlabel
\newcommand{\@currentlabel}{\thetableone}
\makeatother
\renewcommand{\thetableone}{\alph{tableone}}

\tablenotetext{\thetableone}{\label{1mass} Mean stellar mass of galaxies in the corresponding spectroscopic subsample. \stepcounter{tableone}}

\tablenotetext{\thetableone}{\label{1zmean} Mean redshift of galaxies in the corresponding spectroscopic subsample. \stepcounter{tableone}}

\tablenotetext{\thetableone}{\label{1zmin} Lower redshift limit used. \stepcounter{tableone}}

\tablenotetext{\thetableone}{\label{1zmax} Upper redshift limit used. \stepcounter{tableone}}

\tablenotetext{\thetableone}{\label{1velocity} We show adopted velocity dispersion $\sigma_{v}$ for the red galaxies and adopted circular velocity $V_{c}$ for the blue galaxies. \stepcounter{tableone}}

\tablenotetext{\thetableone}{\label{1recover} We show the angular separations of $r_p = 37.6 \hKpc$ and $r_p = 59.6 \hKpc$ which are the separations of the third and fouth data points in Figures~\ref{fig:nwpredWITHred} through \ref{fig:nwpblueWITHblue}. \stepcounter{tableone}}

\end{deluxetable}
\end{center}

\newpage
\clearpage
\begin{center}
\begin{deluxetable}{ccccc}
\tablecolumns{5}
\tablecaption{Fractional mass growth measurements\label{tab:data1}}
\tablehead{
\colhead{Subsample} & $\di (\ln \, M) \, / \, \di \, t \mid_{BT}$\tablenotemark{\protect{\ref{2df}}} & \colhead{blue fraction}\tablenotemark{\protect{\ref{2bluedf}}} & $\di (\ln \, M) \, / \, \di \, t \mid_{KW}$\tablenotemark{\protect{\ref{2max}}} & \colhead{blue fraction}\tablenotemark{\protect{\ref{2bluemax}}} \\
\colhead{} & [$10^{-3}\,\hhGyr$] & [\percent]
           & [$10^{-3}\,\hhGyr$] & [\percent]
}
\startdata

red0   & $               $  &  $    $ & $         $ & $    $\\
red1   & $ 7.9 \pm 4.2$ \tablenotemark{\protect{\ref{2error}}} &  $27.3$ & $13.0 \pm 6.8$ & $31.5$\\
red2   & $ 9.1 \pm 4.8$  &  $28.3$ & $14.6 \pm 7.6 $ & $31.4$\\
red3   & $10.2 \pm 5.3$  &  $25.3$ & $16.3 \pm 8.5 $ & $29.3$\\
red4   & $11.8 \pm 6.1$  &  $21.9$ & $18.9 \pm 9.7 $ & $26.3$\\
red5   & $12.5 \pm 6.4$  &  $23.6$ & $20.3 \pm 10.4$ & $27.7$\\
red6   & $13.0 \pm 6.7$  &  $20.9$ & $21.2 \pm 10.9$ & $25.6$\\
red7   & $14.3 \pm 7.3$  &  $20.7$ & $23.6 \pm 12.1$ & $25.3$\\
red8   & $15.2 \pm 7.8$  &  $19.1$ & $26.1 \pm 13.3$ & $23.5$\\
red9   & $18.2 \pm 9.3$  &  $11.8$ & $37.4 \pm 19.0$ & $15.7$\\
blue0  & $            $  &  $    $ & $         $ & $    $\\
blue1  & $ 3.2 \pm 1.7$  &  $48.3$ & $6.0  \pm 3.1 $ & $46.5$\\
blue2  & $ 3.6 \pm 1.9$  &  $47.3$ & $6.6  \pm 3.4 $ & $47.0$\\
blue3  & $ 3.8 \pm 1.9$  &  $44.1$ & $6.8  \pm 3.5 $ & $45.1$\\
blue4  & $ 3.9 \pm 2.0$  &  $40.7$ & $6.9  \pm 3.5 $ & $43.2$\\
blue5  & $ 4.3 \pm 2.2$  &  $42.4$ & $7.2  \pm 3.8 $ & $45.8$\\
blue6  & $ 4.6 \pm 2.4$  &  $36.0$ & $7.7  \pm 4.0 $ & $41.6$\\
blue7  & $ 4.8 \pm 2.6$  &  $36.8$ & $8.3  \pm 4.4 $ & $42.2$\\
blue8  & $ 5.5 \pm 2.9$  &  $27.9$ & $9.5  \pm 5.0 $ & $34.4$\\
blue9  & $ 5.3 \pm 2.9$  &  $23.0$ & $10.5 \pm 5.7 $ & $28.4$\\
\enddata

\tablecomments{Fractional mass growth of Main Galaxy by merging $\hhGyr$, split by spectroscopic subsample.}

\setcounter{tabletwo}{1} \makeatletter \let\@currentlabel\oldlabel
\newcommand{\@currentlabel}{\thetabletwo} \makeatother
\renewcommand{\thetabletwo}{\alph{tabletwo}}

\tablenotetext{\thetabletwo}{\label{2df} Measurements under the 
assumption $t_{merge,i} = t_{BT,i}$. \stepcounter{tabletwo}}

\tablenotetext{\thetabletwo}{\label{2bluedf} Percent contribution of the blue galaxies from the imaging sample under the assumption $t_{merge,i} = t_{BT,i}$. \stepcounter{tabletwo}}

\tablenotetext{\thetabletwo}{\label{2max} Measurements under the assumption $t_{merge,i} = t_{KW,i}$. \stepcounter{tabletwo}}

\tablenotetext{\thetabletwo}{\label{2bluemax} Percent contribution of the blue galaxies from the imaging sample under the 
assumption $t_{merge,i} = t_{KW,i}$. \stepcounter{tabletwo}}

\tablenotetext{\thetabletwo}{\label{2error} Errors are estimated including the error from merger time. \stepcounter{tabletwo}}

\end{deluxetable}
\end{center}

\newpage
\clearpage
\begin{center}
\begin{deluxetable}{ccc}
  \tablecolumns{3}
  \tabletypesize{\footnotesize}
  \tablewidth{0pt}
\tablecaption{Fractional mass growth measurements for massive red galaxies\label{tab:red}}
\tablehead{
\colhead{Subsample} 
& $\di (\ln \, M) \, / \, \di \, t \mid_{BT}$\tablenotemark{\protect{\ref{3df}}} 
& $\di (\ln \, M) \, / \, \di \, t \mid_{KW}$\tablenotemark{\protect{\ref{3max}}} \\
\colhead{} & [$10^{-3}\,\hhGyr$]
           & [$10^{-3}\,\hhGyr$]
}
\startdata
                & $                  $  &  $                  $ \\
red7            & $  31.8 \pm 15.9  \tablenotemark{\protect{\ref{3error}}}  $  &  $    52.6 \pm 26.3 $ \\
red8            & $  33.8 \pm 16.9   $  &  $    58.1 \pm 29.1 $ \\
red9            & $  40.5 \pm 20.3   $  &  $    83.1 \pm 41.6 $ \\
                & $                  $  &  $                  $ \\
$\gtrsim L^*$   & $                  $  &  $                  $ \\
red7$\sim$9     & $   37.0 \pm  18.5 $  &  $   70.5 \pm  35.3 $ \\
                & $                  $  &  $                  $ \\
$\geqslant L^*$ & $                  $  &  $                  $ \\
red8$\sim$9     & $   38.3 \pm  19.1 $  &  $   74.9 \pm  37.4 $ \\
                & $                  $  &  $                  $ \\
LRG             & $                  $  &  $                  $ \\
red9            & $  40.5 \pm 20.3   $  &  $    83.1 \pm 41.6 $ \\
\enddata

\tablecomments{Fractional mass growth measurements for massive red galaxies
($L^*$ galaxies and above)
under assumption of merger rates $\propto (1+z)^{+3.0 \pm 1.1}$, split by spectroscopic subsample.}

\setcounter{tablethree}{1} \makeatletter \let\@currentlabel\oldlabel
\newcommand{\@currentlabel}{\thetablethree} \makeatother
\renewcommand{\thetablethree}{\alph{tablethree}}

\tablenotetext{\thetablethree}{\label{3df} Measurements under the 
assumption $t_{merge,i} = t_{BT,i}$. \stepcounter{tablethree}}

\tablenotetext{\thetablethree}{\label{3max} Measurements under the assumption $t_{merge,i} = t_{KW,i}$. \stepcounter{tablethree}}

\tablenotetext{\thetablethree}{\label{3error} Errors are estimated including the error from merger time. \stepcounter{tablethree}}

\end{deluxetable}
\end{center}

   \clearpage

   \clearpage

   \clearpage

   \clearpage

   \clearpage

   \clearpage

   \clearpage

   \clearpage

   \clearpage

   \clearpage

   \clearpage

   \clearpage

   \clearpage

   \clearpage

   \clearpage

   \clearpage

   \clearpage

   \clearpage

   \clearpage

   \clearpage

   \clearpage

   \clearpage

   \clearpage

   \clearpage

   \clearpage

   \clearpage

\end{document}